\documentclass[aps,prb,twocolumn]{revtex4}
\usepackage{amssymb}
\usepackage{amsmath}
\usepackage{amsbsy}
\usepackage{graphicx,epsfig,psfrag}
\newcommand{\w}{\omega}
\newcommand{\beq}{\begin{equation}}
\newcommand{\eeq}{\end{equation}}
\newcommand{\bea}{\begin{eqnarray}}
\newcommand{\eea}{\end{eqnarray}}
\newcommand{\beas}{\begin{eqnarray*}}
\newcommand{\eeas}{\end{eqnarray*}}
\newcommand{\nn}{\nonumber}

\newcommand{\pdag}{{\phantom{\dagger}}}

\renewcommand{\vec}[1]{{\mathbf #1}}

\begin{document}

\title{
Quantum phase transition of Ising-coupled Kondo impurities
}
\author{M. Garst$^a$, S. Kehrein$^b$, T. Pruschke$^b$, A. Rosch$^a$, and M. Vojta$^a$}
\affiliation{$^a$Institut f\"ur Theorie der Kondensierten Materie,
Universit\"at Karlsruhe, Postfach 6980, 76128 Karlsruhe, Germany\\
$^b$Theoretische Physik III, Elektronische Korrelationen und
Magnetismus, Universit\"at Augsburg, 86135 Augsburg, Germany}
\date{February 10, 2004}
\begin{abstract}
  We investigate a model of two Kondo impurities coupled via an Ising
  interaction.  Exploiting the mapping to a generalized
  single-impurity Anderson model, we establish that the model has a
  singlet and a (pseudospin) doublet phase separated by a
  Kosterlitz--Thouless quantum phase transition.
  Based on a strong-coupling analysis and renormalization group arguments,
  we show that at this transition the conductance $G$ through the system either
  displays a zero-bias anomaly, $G\sim |V|^{-2(\sqrt{2}-1)}$,
  or
  takes a universal value, $G=\frac{e^2}{ \pi \hbar}
  \cos^2\!\left(\frac{\pi}{2 \sqrt{2}}\right)$,
  depending on the experimental setup.
  Close to the Toulouse point of the individual Kondo impurities, the
  strong-coupling analysis allows to obtain the
  location of the phase boundary analytically. For general model
  parameters, we determine the phase diagram and investigate the
  thermodynamics using numerical renormalization group calculations.
  In the singlet phase close to the quantum phase transtion, the
  entropy is quenched in two steps: first the two Ising-coupled spins
  form a magnetic mini-domain which is, in a second step, screened by
  a Kondoesque collective resonance in an effective {\it solitonic}
  Fermi sea.
  In addition, we present a flow equation analysis which
  provides a different mapping of the two-impurity model to a
  generalized single-impurity Anderson model in terms of fully
  renormalized couplings,
  which is applicable for the whole range of model parameters.
\end{abstract}
\pacs{75.20.Hr, 73.21.La, 71.10.Hf}
\maketitle

\section{Introduction}

Kondo physics plays a fundamental role for the low-temperature behavior
of a large variety of physical systems like magnetic impurities in
metals, heavy fermion systems, glasses, quantum dots, etc.
Its key feature is the quenching of the impurity entropy through non-perturbative
screening by many-particle excitations in the associated quantum bath.
For magnetic impurities in metals this amounts to the formation
of the Kondo singlet between the localized spin and electron--hole
excitations in the Fermi sea.~\cite{Hewson}

Most of the aspects of single-impurity Kondo physics are now
well understood after theoretical tools have been developed
that can deal with its intrinsic strong-coupling nature.
\cite{Wilson75,Betheansatz,Hewson}
However, in many physical systems the interaction of different
impurities, i.e., multi-impurity Kondo physics, is important.
For example in heavy fermion systems the RKKY interaction
between different impurity spins leads to
competition between local Kondo physics and long-range magnetic
order that determines their phase diagram.\cite{doniach}
More recently, related questions about coupled two-level systems have
gained much interest in quantum computation, where decoherence due
to unwanted couplings among qubits and between qubits and environment
should be avoided;
on the other hand the intentional coupling of qubits
is the key step to performing quantum logic operations.

In the present paper we investigate the case of two spin-$1/2$ Kondo
impurities $\vec{S}_1,\vec{S}_2$ coupled via a Ising coupling,
\beq
H_{12}^{\rm Ising}=K_z S_1^z S_2^z.
\label{IsingCoupling}
\eeq
The Kondo coupling of each impurity to its bath is given by
\begin{equation}
H_{j}^{K}= 2 J_\perp ( S_j^x s_{0,j}^x + S_j^y s_{0,j}^y)  + 2 J_z S_j^z s_{0,j}^z
\label{hintro}
\end{equation}
where $j=1,2$ labels the impurity, and $\vec{s}_{0,j}$ is the bath spin
operator at the respective impurity site.
Furthermore the two baths are disconnected.

Two impurities coupled both to baths and among each other
present the simplest realization of the so-called cluster Kondo effect,
which has been discussed, e.g., in context of disordered Kondo--lattice
compounds.\cite{antonio}
Furthermore, the dynamics of magnetic droplets or domains, formed in
disordered itinerant systems near a magnetic quantum phase
transition,\cite{millis}
also leads to models of coupled impurities like the one considered
here.
Therefore, we will also refer to the two coupled impurities as
magnetic ``mini-domain''.

In the context of Kondo impurities, an Ising-like
coupling (\ref{IsingCoupling})
can be thought of as an effective impurity interaction for
heavy fermion systems with an easy axis.
Also, Ising coupling appears naturally in quantum dots
that are coupled via their mutual capacitance \cite{Waugh95}
here the two-level
systems are pseudospins representing the number of electrons on the
dots, and therefore SU(2) symmetry is broken from the
outset.\cite{Matveev91,Matveev95,Golden96}
Equivalently, one can think of two two-level systems with transversal
coupling, with the experimental realization of coupled flux
qubits.\cite{Mooij99}
We will discuss different formulations and applications of our model
in the body of the paper.

\begin{figure}
\centerline{
\psfrag{Jz}{$J_z$}
\psfrag{Jzc}{$J^{\rm cr}_z$}
\psfrag{1}{$\mathcal{O}(1)$}
\psfrag{0}{$0$}
\psfrag{Tk}{$\displaystyle \frac{T_K^{(1)}}{K_z}$}
\includegraphics[width=3in]{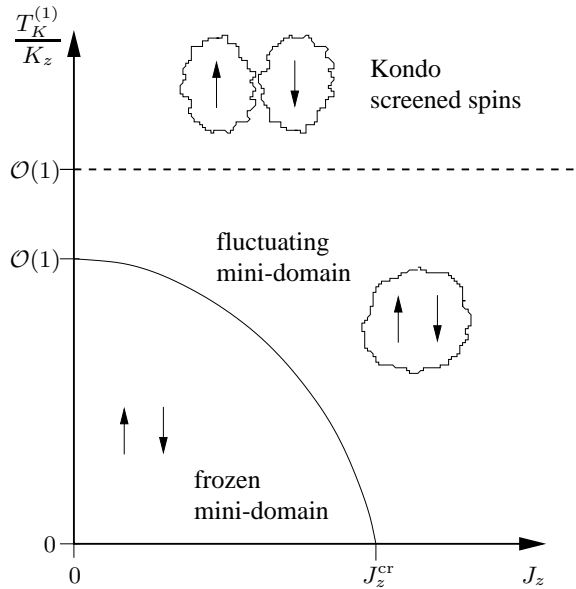}
}
\caption{
  Schematic phase diagram for two Ising-coupled Kondo impurities
  (\ref{IsingCoupling},\ref{hintro}).
  The vertical axis denotes the ratio between
  single-impurity Kondo temperature $T_K^{(1)}$ (determined by
  the Kondo couplings $J_z$, $J_\perp$, and the bath bandwidth $D$)
  and Ising coupling $K_z$.
  The horizontal axis ($J_z$) measures the anisotropy of the Kondo
  coupling; in the universal regime, $T_K^{(1)} \ll D$, isotropic Kondo
  coupling corresponds to $J_z \ll J_z^{cr}$.  The model has two
  phases: At small $J_z$ and large $K_z$, the ground state is doubly
  degenerate and the two impurity spins are locked into a ``frozen
  mini-domain''.  In contrast, at large $J_z$ or small $K_z$ the
  ground state is a singlet with local Fermi-liquid characteristics.
  The quantum phase transition is of Kosterlitz--Thouless type.
  In the universal regime, $T_K^{(1)} \ll D$ implying $J_\perp \ll D$,
  $T_K^{(1)}$ is the only low-energy scale of the single-impurity
  problem, and the phase transition occurs at a critical $K_z$
  proportional to $T_K^{(1)}$ with the proportionality factor
  depending on $J_z$, as shown in the figure (solid line).
  The critical $K_z$ diverges as $J_z\to J_z^{cr}$.
  Below the dashed crossover line the two spins form an Ising mini-domain:
  the low-energy fluctuations are associated with the pseudospin degree
  of freedom, i.e., for antiferromagnetic $K_z$ the staggered impurity
  susceptibility is much larger than the uniform one.
}
\label{fig:phd1}
\end{figure}

Coupled impurities or two-level systems have been investigated in a
number of papers,\cite{2imp,2impnrg,2impsakai,2impfye,2impcft,2impPhase,2impbos,2impoli} where
most attention has been focussed on the case of SU(2)-symmetric direct
exchange coupling between the impurity spins,
$K\,\vec S_1\cdot\vec S_2$.
Here, two different regimes are possible as function
of the inter-impurity exchange~$K$:
for large antiferromagnetic $K$ the impurities combine to
a singlet, and the interaction with the conduction
band is weak, whereas for ferromagnetic $K$ the impurity
spins add up and are Kondo-screened by conduction electrons in the low-temperature limit.
Notably, there is
{\em no} quantum phase transition as $K$ is varied
in the generic situation without particle--hole symmetry
(whereas one finds an unstable non-Fermi liquid fixed
point in the particle--hole symmetric
case).\cite{2impnrg,2impsakai,2impcft}

As has been pointed out
by Andrei {\it et al.},\cite{Andrei} the case of Ising coupling is
different and particularly interesting, because
for large $|K_z|$ the two Ising-coupled spins form a magnetic
mini-domain which still contains an internal degree of freedom as the
ground state of $H^{\text{Ising}}_{12}$ is doubly degenerate
(in contrast to the inter-impurity singlet mentioned above).
For the case of antiferromagnetic $K_z$ (which we will assume in the following),
the two low-energy states of the impurities (forming a pseudospin)
are $|\!\uparrow\downarrow\rangle$ and $|\!\downarrow\uparrow\rangle$.
As we will show in this paper, the fate of this pseudospin degree of freedom
depends on the strength and asymmetry of the Kondo coupling $J$
between the spins and the bath electrons.

\subsection{Summary of results}

Here, we will summarize our main results which are detailed in the
body of the paper, and schematically represented in the phase diagram
of Fig.~\ref{fig:phd1}.
A brief summary of the  methods used to obtain these results is given
below in Sec.~\ref{summary.methods}.

The model of two Ising-coupled impurities, connected to two {\em
separate} fermionic reservoirs (realized, e.g., by attaching two
separate leads to two quantum dots, Fig.~\ref{fig:model}), has two
ground-state phases associated to either a screened or an unscreened
pseudospin.  For small Kondo couplings $J_\perp$, $J_z$ and large
$K_z$, tunneling between the two pseudospin configurations,
$|\!\uparrow\downarrow\rangle$ and $|\!\downarrow\uparrow\rangle$, is
supressed at low energies, i.e., the mini-domain is ``frozen'' as $T\to
0$, and the ground state entropy is $S_0=\ln 2$.
In contrast, for small $K_z$ the two impurities are individually
Kondo screened,
resulting in a Fermi-liquid phase with vanishing residual entropy.
This implies the existence of a quantum phase transition for $K_z \sim
T_K^{(1)}$, where $T_K^{(1)}$ is the single-impurity Kondo temperature.
For isotropic Kondo coupling, i.e., small $J_z$, this has been previously
pointed out by Andrei {\it et al.}\cite{Andrei}

What is the nature of this phase transition? Does it occur by breaking
up the Ising-coupled mini-domain, or rather by strong fluctuations of
the preformed pseudospin? What are the universal properties of this
transition? The key observation, which helps to answer these
questions, is that the system can also be tuned towards the quantum
phase transition by increasing the Ising component, $J_z$, of the
coupling of the spins to the environment in a regime where $K_z \gg
T_K^{(1)}$ (see Fig.~\ref{fig:phd1}).
For $K_z \gg T_K^{(1)}$ the
mini-domain is stable. However, upon increasing $J_z$ a
many-particle effect (related to formation of a Mahan
exciton~\cite{mahan}) enhances the tunneling between the two
configurations, $|\!\uparrow\downarrow\rangle$ and
$|\!\downarrow\uparrow\rangle$, of the mini-domain.  If $J_z$ exceeds
a critical value, $J_z^{cr}$, this tunneling can quench the pseudospin
even for infinitesimal $T_K^{(1)}$ (equivalent to infinitesimal
transverse Kondo coupling $J_\perp$).

The resulting phase of the ``fluctuating mini-domain'' is actually a
Fermi liquid with vanishing residual entropy.  Note that the
finite-temperature properties in this regime are rather different from
that of the Fermi liquid which is obtained for $T_K^{(1)} \gg K_z$.  For
large $J_z$ and small $J_\perp$, the high-temperature $\ln 4$ impurity
entropy is quenched in two stages: first, at the scale $T^0 \approx
K_z$, the mini-domain ``forms'', quenching half of entropy; second
the strong fluctuations kill the remaining $\ln 2$ entropy at a much
lower scale $T^\ast$, this scale $T^\ast$ can be identified with a
collective Kondo temperature associated to pseudospin screening.
(Note that this type of two-stage screening is completely different
from the one occurring for two conventional Kondo screening channels
with different strengths.\cite{2imp})  In contrast, for $T_K^{(1)} \gg K_z$
the entropy of the two spins is quenched simultaneously in a single
step at the scale $T_K^{(1)}$.  Nevertheless, the two Fermi-liquid
regimes are adiabatically connected by a smooth crossover, which we
will show to be identical to the well-known crossover in the
single-impurity Anderson model from the mixed valence into the Kondo
regime.

The quantum phase transition at $J_z=J_z^{cr}$ turns out to be in the
Kosterlitz--Thouless universality class. Assuming continuity along the
phase boundary (which we verified numerically), this is true for the
transition at arbitrary $J_z\le J_z^{cr}$.
Furthermore, universality immediately implies that
the ``fluctuating mini-domain'' regime with its characteristic
two-stage quenching of the entropy also exists for small $J_z$ close
to the quantum phase transition (see Fig.~\ref{fig:phd1}).

Physical observables, like the conductance in a quantum dot setup,
show {\em universal} behavior in the vicinity of the phase boundary.
Therefore, we can calculate them close to $J^{cr}_z$ (see
Fig.~\ref{fig:phd1}), where the phase transition takes place in a
regime of large $K_z$ being accessible to a strong-coupling
analysis combined with renormalization group arguments.
Depending on the experimetal setup, we find, e.g.,
a conductance anomaly characterized by the exponent $-2(\sqrt{2}-1)$,
or
a universal conductance $e^2 \cos^2\left[\frac{\pi}{2 \sqrt{2}}\right]/(\hbar \pi)$ at the phase
transition.
We emphasize again that these results are valid close to the quantum phase
transition even for small $J_z$ where a strong-coupling analysis is not possible.

\subsection{Methods and outline}
\label{summary.methods}

To obtain the physical picture and the results described above,
we use a combination of six different and partly complementary
methods.

In Sec.~\ref{sec:Model} we (i) map our model of Ising-coupled spins
to a generalized Anderson model by bosonization and refermionization
techniques.
This mapping is used to obtain the qualitative
structure of the phase diagram and analytic results for the phase
boundary at large $J_z$.
Furthermore, for the generalized Anderson model it is much easier to
implement (ii) numerical renormalization group (NRG), which is
presented in Sec.~\ref{sec:NRG}.
With the help of NRG it is possible to determine
numerically the phase boundaries in regimes not accessible to analytic
methods. Furthermore, NRG is essential to establish that the phase
transition at small $J_z$ is continuously connected to the one at
large $J_z$.

Making use of this adiabatic continuity is the main idea of this paper
to obtain analytic results for the quantum phase transtion. By
increasing $J_z$ we can tune the transition from a regime with
$K_z\sim T_K^{(1)}$ to a regime with $K_z \gg T_K^{(1)}$, where we can
employ (iii) a strong-coupling expansion
(Sec.~\ref{sec:StrongCoupling}). The strong-coupling result is
analyzed using (iv) perturbative renormalization group, or more
precisely power counting, taking into account the anomalous dimensions
created from an orthogonality catastrophe.
Using these methods the
phase diagram for large $K_z\gg T_K^{(1)}$ and the precise position of
the critical point for $K_z \to \infty$ can be obtained.
In addition, we can determine the relevant phase shifts and scaling dimensions
of leading relevant and irrelevant operators (Sec.~\ref{sec:Pert}).
This allows us to analytically calculate the conductance and zero-bias anomalies
close to the quantum phase transition for {\em arbitrary} values
of $ T_K^{(1)}/K_z$, see Sec.~\ref{sec:Appl}.
To analytically obtain the precise shape of the phase diagram for large $J_z$,
we develop in App.~\ref{sect:SW}
(v) a generalization of the Schrieffer--Wolff transformation to take into
account the short-range density interaction of our generalized Anderson model,
leading to associated power-law singularities.

In Sec.~\ref{sec:Flow} and App.~\ref{sect:App_feq} we re-derive some above
results independently by using (vi) flow equations.
The advantage of this method is that it has a broad range of applicability
and gives a more natural description in terms of renormalized quantities.
The flow equation mapping nicely establishes the
equivalence of the different Fermi-liquid regimes of the model.
It also allows us to derive the full phase diagram
analytically for general values of the coupling $J_z$ that are
not accessible with the strong-coupling expansion.

Transport quantities and corresponding possible experimental
realizations are discussed in Sec.~\ref{sec:Appl}.
The most promising way to implement Ising-coupled (pseudo)spin variables
is the use of charge degrees of freedom of two quantum dots,
following Matveev's proposal \cite{Matveev91,Matveev95},
and employ a capacitive coupling which takes the role of $K_z$,
see Sec.~\ref{capacity}.

Most of our methods are based on the mapping of the original two-impurity model
to a generalized single-impurity Anderson model, except for the
strong-coupling analysis in Sec.~\ref{sec:hstrong} which is applied
to the original model (and thus directly establishes the existence of a
phase transition).
As we will show, the employed mapping provides a particularly clear picture of
the underlying physics, e.g., it establishes the universality class of the
transition, and allows to make further progress using flow equations.
Thus, the mapping turns out to be extremely helpful for obtaining the complete
picture presented below.


\section{Model: Variations and transformations}
\label{sec:Model}

In this section we discuss the various formulations
of the model under consideration, together with the mapping between
them, which is based on the well-known relation between the
spin-boson model and the anisotropic Kondo model.\cite{Betheansatz,Weiss,RMP}

Throughout this paper, we will consider the so-called scaling limit
where both $K_z$ and the single-impurity Kondo temperature $T_K^{(1)}$
are much smaller than the high-energy cutoff $D$ of the theory,
$T_K^{(1)},K_z \ll D$.
Keeping $J_z\geq 0$ and $D$ fixed, this implies $J_\perp\to 0$.
Only in this scaling limit the models discussed
below can be mapped upon each other.
In general, the position of the phase boundary, i.e., the value of the
critical coupling $K_z^{cr}$ depends on microscopic details. In the
scaling limit, however, $K_z^{cr} / T_K^{(1)}$ depends only on $J_z$
which parametrizes the renormalization flow in the single-impurity
model, see Sec.~\ref{sec:NRG:boundary}.  This universality is
represented in the phase diagram in Fig.~\ref{fig:phd1}.

\begin{figure}
\centerline{
\psfrag{J}{$\vec{J}$}
\psfrag{K}{$K_z$}
\includegraphics[width=3.2in]{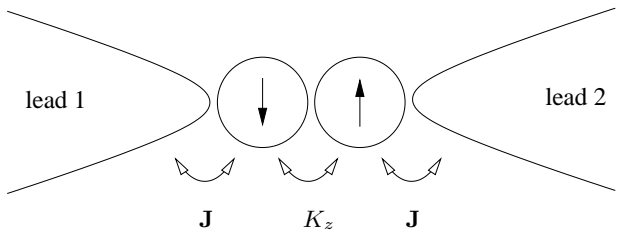}
}
\caption{
  Schematic plot of a system represented by the Hamiltonian (\ref{Model}).
  Two spins are coupled via an (anisotropic) exchange interaction $J$ to two leads.
  The spins interact via an Ising coupling  $K_z$.
  For an explicit discussion of possible experimental setups see Sec.~\protect\ref{sec:Appl}.
}
\label{fig:model}
\end{figure}
\subsection{Ising-coupled Kondo impurities}

We consider the model
\bea \label{Model}
H^K=H_1^K+H_2^K+H^{\rm Ising}_{12}\,,
\eea
where two spins $\vec{S}_1$ and $\vec{S}_{2}$ interact by the Ising
interaction (\ref{IsingCoupling}), $H^{\rm Ising}_{12}$.
Each of the spins couples to a separate fermionic bath,
$c^\pdag_{k \alpha j}$, via
an anisotropic Kondo Hamiltonian $H_j^K$ ($j=1,2$), see Fig.~\ref{fig:model},
\begin{equation}
\label{Kondo}
H_{j}^{\rm K} = H_0[\Psi_{\alpha,j}]
+ \sum_{n\alpha\beta} J_{n} S_{j}^n
\Psi^\dag_{\alpha j}(0) \sigma^n_{\alpha\beta} \Psi^\pdag_{\beta j}(0)
\end{equation}
where $\alpha,\beta$ are spin indices and
$H_0[\Psi_{\alpha,j}] =\sum_{k, \alpha} \epsilon_k  c^\dag_{k \alpha j} c^\pdag_{k \alpha j}$
with $\Psi_{\sigma j}(x) = \sum_k {\rm e}^{- i k x} c_{k \sigma j}$.
The exchange
coupling is assumed to be the same for both impurities and has an
anisotropic form, $J_{n} = (J_\perp,J_\perp,J_z)$.

The fermionic baths are assumed to be particle--hole symmetric
with a bandwidth $D$; for a rectangular band the density of states
at the Fermi level is then $\rho_F = 1/D$.
Our conclusions are not modified by the presence of a particle--hole
asymmetry.
A comprehensive discussion of possible modifications of our Hamiltonian
(\ref{Model}), e.g., due to tunneling between the fermionic baths,
and how they influence our results, will be given
in Sec.~\ref{sec:Pert}.

A model of form (\ref{Model}) may be approximately realized with
real spins in the presence of a strong Ising anisotropy.
In addition, it occurs naturally as a model for capacitively
coupled quantum dots,\cite{Andrei,Matveev91,Matveev95,Golden96}
where the local operators $S_1^z$ and $S_2^z$ describe charge states, i.e.,
pseudospin degrees of freedom, on the two dots.
Concrete application of our results to such a situation will
be discussed in Sec.~\ref{sec:Appl}.

\subsection{Coupled Qubits in ohmic baths}

An alternative starting point for our model of Ising-coupled impurities
can be formulated in terms
of two two-level systems (spin-boson models),
$H^{\rm SB}=H_1^{\rm SB}+H_2^{\rm SB}+H_{12}^{\rm trans}$
with
\begin{equation}
\label{HSB}
H_j^{\rm SB}= H_0[b_{k j}] + \frac{\Delta}{2}\,\sigma_j^x
+\frac{1}{2}\sum_{k>0} \lambda^\pdag_{k}\,\sigma_j^z(b^\pdag_{k j}+b^\dag_{k j})
\,,
\end{equation}
where $H_0[b_{k j}] = \sum_{k>0} \omega^\pdag_{k}\,b^\dag_{k j} b^\pdag_{k j}$,
and a transversal coupling between them,
\begin{equation}
H_{12}^{\rm trans} =
\frac{K_z}{4}\:\sigma_1^z\,\sigma_2^z \,.
\end{equation}
Here $b^\dag_{k j}$
are the bosonic creation operators for heat bath~\#$j$.  $\Delta$ is
the bare tunneling matrix element between the two levels. The impurity
properties are completely parametrized by the spectral function
$J(\omega)\equiv\sum_k
\lambda_{k}^2\,\delta(\omega-\omega_{k})$, which we assume to be of
Ohmic form, $J(\omega)=2\alpha\,\omega\,e^{-\omega/2\omega_c}$.

One realization of this model is the interaction of tunneling centers
in glasses through higher-order phonon exchange.\cite{Kassner90}
In the context of quantum computation this model arises in studies of
decoherence of coupled superconducting qubits:
the transversal coupling $K_z$ is generated through a superconducting
flux transporter, and the heat baths describe the environment leading to
decoherence.\cite{Mooij99,Wilhelm02}
Here, the assumption of two different baths for the two qubits is
justified, e.g., if the baths model electromagnetic noise coming
from read-out circuits, which are separate for each qubit.

\subsection{Bosonization}
\label{sec:Bosonization}

The equivalence of $H^{\rm K}$ and $H^{\rm SB}$ can be explicitly
shown in the framework of bosonization. Furthermore, we will demonstrate
that both Hamiltonians can be mapped to a generalized Anderson impurity
model. We will use this mapping extensively, both to solve the models
numerically within NRG and to identify the position and nature of the
quantum phase transition in certain limits analytically.

It is well known\cite{Weiss} that both the Kondo model and the spin-boson
model are equivalent to a generalized resonant-level model to
be defined below.
Our model, $H^{\rm K}$ and $H^{\rm SB}$, however, consists
of two coupled Kondo- and spin-boson Hamiltonians. The
crucial point is that the assumed coupling $H^{\rm Ising}_{12}$ and
$H^{\rm trans}_{12}$, respectively, is transformed trivially by
switching between these three representations.

We start from the two Kondo Hamiltonians (\ref{Kondo}) and apply the
bosonization identity
\bea \label{boso}
\Psi_{\sigma j}(x) = \frac{1}{\sqrt{2 \pi a}} \, F_{\sigma j}\; {\rm e}^{-i \phi_{\sigma j}(x)}\,,
\eea
where $a$ is a short distance cutoff, $F_{\sigma j}$ is an
anticommuting Klein factor ($\{F_{\sigma j}^\dag,F_{\sigma'j'}^\pdag\}=
2\delta_{jj'}\delta_{\sigma\sigma'}$),
and $\phi_{\sigma j}$ is the corresponding bosonic
field with $[\phi_{\sigma j}(x),\partial_{x'} \phi_{\sigma' j'}(x')]=2 \pi
i \delta(x-x') \delta_{jj'} \delta_{\sigma\sigma'}$.
Transforming to bosonic charge and spin fields,
$
\phi_{s/c, j} = \frac{1}{\sqrt{2}} \left( \phi_{\uparrow j} \pm \phi_{\downarrow j}\right),
$
the bosonized version  of the Kondo Hamiltonians (\ref{Kondo}) reads
\bea
H_j ^K&=&
H_0[\phi_{c j}] + H_0[\phi_{s j}] + \frac{J_z}{\sqrt{2} \pi} S^z_j \partial_x \phi_{s j}(0) \\
&&
+ \frac{J_\perp}{2 \pi a} \left({\rm e}^{-i \sqrt{2} \phi_{s j}(0)} S^+_{j} F^\dag_{\downarrow j} F^\pdag_{\uparrow j}
+ {\rm h.c.} \right)\,,
\nonumber
\eea
where $H_0[\phi]=v_F \int \frac{{\rm d}x}{2 \pi} \frac{1}{2} (\partial_x
\phi)^2$ assuming a linear dispersion, $\epsilon_k=v_F k$.  The bosonic
charge field, $\phi_{c j}$, decouples and is omitted in the following.

Applying a general Emery--Kivelson transformation, \cite{RMP}
\bea
U_\gamma=\exp\left(i \gamma \sum_{j} S_{z j} \phi_{s j}(0) \right) \,,
\label{EK}
\eea
parametrized by $\gamma$,
the Hamiltonian $H_j^K$ transforms into $\tilde{H}_j ^K=U_\gamma H_j^K
U_\gamma^\dagger$,
\bea
\label{Htilde}
\tilde{H}_j ^K&=&  H_0[\phi_{s j}] +\left(\frac{J_z}{\sqrt{2} \pi} - \gamma v_{\rm F}\right) S^z_j \partial_x \phi_{s j}(0)
\\
&&
+ \frac{J_\perp}{2 \pi a} \left({\rm e}^{-i \left(\sqrt{2} - \gamma\right) \phi_{s j}(0)} S^+_{j}
F^\dag_{\downarrow j} F^\pdag_{\uparrow j}
+ {\rm h.c.} \right)\nonumber \,.
\eea
Importantly, the Ising coupling
(\ref{IsingCoupling}) is {\em not} affected by this transformation,
$H_{12}^{\text{Ising}}=U_\gamma H_{12}^{\text{Ising}}
U_\gamma^\dagger$.

For two special values of the transformation parameter $\gamma$
the Emery--Kivelson transformation results in particularly
interesting forms of the Hamiltonian.
First consider the case when $\gamma = \sqrt{2}$.
The exponents in the spin flip term of (\ref{Htilde}) then vanish
and $\tilde{H}^K_j$ can be cast into the form of the spin-boson Hamiltonian (\ref{HSB}).
We can now easily identify the coupling constants, $\Delta = \frac{J_\perp}{\pi a}$, $\lambda_k
=\left(\frac{J_z}{\sqrt{2} \pi} - \sqrt{2} v_{\rm F}\right)
\sqrt{\frac{2\pi k}{L}}$ ,  $ \omega_{k}=v_{\rm F} k$
and $b_{q j}$ are the Fourier components of $\partial_x \phi_{s j}(x)$
with $\phi_{s j}(x) = - \sum_{k>0} \sqrt{\frac{2 \pi}{k L}} \left( -i
  b^\pdag_{k j}{\rm e}^{- i k x} + i b^\dag_{k j}{\rm e}^{i k
    x}\right) {\rm e}^{- k a/2}$. The linear dispersion and the form
of the coupling $\lambda_q$ result in a ohmic form of the spectral
function $J(\omega)=2\alpha\,\omega\,e^{-\omega/2\omega_c}$
with a strength
\begin{equation}
\alpha = (J_z \rho - 1)^2 \,,
\end{equation}
where $\rho$ is the density of states,  $\rho = 1/(2\pi v_{\rm F})$.

The single-impurity Kondo temperature has in general a power law
dependence on the ``tunneling rate'' $J_\perp$,
\begin{equation}
T_K^{(1)} \propto J_\perp^\frac{1}{1-\alpha} ~~~(\mbox{for} J_\perp \ll J_z)\,,
\end{equation}
with $\alpha$ introduced above.

\subsection{Generalized Anderson impurity model}
\label{sec:GAIM}

Applying the Emery--Kivelson transformation with $\gamma = \sqrt{2} - 1$
results in exponentials in (\ref{Htilde}) having the same form as in the
bosonization identity (\ref{boso}) and can therefore be expressed as
fermions $\Psi_{j}$.
The refermionized Hamiltonian can be identified
with a generalized resonant-level model~\cite{Schlottmann78,Delft98}
\bea \label{RL}
H^{\rm RL}_j &=&  H_0[\Psi_{j}]
+ V \left(d^\dag_j  \Psi^\pdag_{j}(0) + {\rm h.c.} \right)
\\
&& + W \left(d^\dag_j d^\pdag_j -\frac{1}{2}\right)  :\Psi^\dag_{j}(0) \Psi^\pdag_{j}(0):
\nonumber
\eea
where $S^z_j=d^\dagger_j d^\pdag_j-\frac{1}{2}$, $V=
\frac{J_\perp}{\sqrt{2 \pi a}}$ and $W = \sqrt{2}J_z - (\sqrt{2} -
1)/\rho$.
$\Psi$ and $d$ are fermionic operators, where
$\Psi$ represents solitonic spin excitations of the original conduction band,
and $d$ describes the spin degree of freedom of the impurity.
The coupling $W$
vanishes for $J_z\rho =1-1/\sqrt{2}$ (or $\alpha=1/2$ for the
spin-boson model), the so-called Toulouse point of the Kondo model;\cite{Toulouse69}
in this case
Eq. (\ref{RL}) reduces to the conventional resonant-level model.
Furthermore, $W<0$ for isotropic small Kondo couplings,
$J_z = J_\perp \ll D$.

In the new variables the Ising interaction takes the form $K_z
(d^\dagger_1 d^\pdag_1-\frac{1}{2})(d^\dagger_2
d^\pdag_2-\frac{1}{2})$. If we interpret the bath index $j=1,2$ as a
pseudospin index $\sigma=\uparrow,\downarrow$, we can identify the
total Hamitonian (\ref{Model}) with a generalized
single-impurity Anderson model
\bea
H^{\rm A} &=& H_0[\Psi_{\sigma}] + V\sum_\sigma
\left(d^\dag_\sigma \Psi^\pdag_{\sigma}(0) + {\rm h.c.} \right) + K_z
\bar{n}_{d\uparrow} \bar{n}_{d\downarrow} \nonumber
\\
&& + W \sum_\sigma \bar{n}_{d\sigma} :\Psi^\dag_{\sigma}(0)
\Psi^\pdag_{\sigma}(0): \label{Anderson}
\eea
with $\bar{n}_{d\sigma}=d^\dag_\sigma
d^\pdag_\sigma-\frac{1}{2}$.
In this representation, the Ising
interaction translates to a local Coulomb repulsion, and $W$ corresponds
to an interaction of the localized level $d_\sigma$ with the
surrounding electrons.
In the limit $K_z=0$ the Hamiltonian describes the extensively studied
x-ray threshold problem.\cite{mahan}
On the other hand, at the Toulouse point where
$W=0$, the standard impurity Anderson Hamiltonian is recovered.
For large $K_z$ the $d$ level is mainly singly occupied; its ``spin''
$\sigma$ corresponds precisely to the pseudospin degree of freedom of
the original mini-domain.  Note that the particle--hole symmetry of
the effective Anderson model corresponds to the symmetry under a rotation
by $\pi$ around the $x$-axis in spin space for the original model.

The mapping of the original two-impurity model onto the
generalized Anderson model (\ref{Anderson}) is one of the central
results of our paper, and will be extensively used in the
numerical study of the phase diagram and the interpretation of the
results.

\subsection{Parameter mapping via phase shifts}

It is important to note that the precise relation of the three models
$H^{SB}$, $H^K$, and $H^A$ depends on the cutoff structure, i.e., on
properties at high energies and short distances.  All formulas quoted
above which relate the various coupling constants are actually only
valid within the cutoff scheme underlying bosonization. However, it is
generally believed that all three models are equivalent independent of
the cutoff structure, as long as one considers only the universal
low-energy properties in a regime where $K_z$ and $T_K^{(1)}$ are much
smaller than any other scale.

We consider now the (non-universal) mapping of model parameters within
different cutoff schemes.
For small values of $J_\perp$ (or $\Delta$ and
$V$) it is possible to calculate the precise mapping by investigating
the perturbation theory in $J_\perp$, $\Delta$, and $V$, respectively
using the fact that all three models map onto a Coulomb
gas~\cite{Yuval70,Weiss,RMP} -- we will not attempt this
here because it is difficult to do it analytically
for an arbitrary cutoff scheme.
However, the mapping of $J_z$, $W$, and $\alpha$ can be obtained
directly by matching the conduction electron phase shifts in the
limit $J_\perp,V=0$, as phase shifts are measurable low-energy
properties.

In the Kondo model (\ref{Kondo}), we denote the scattering phase
shift for antiparallel conduction electron and impurity spins
by $\delta_{J_z}$; for parallel spins the phase
shift is then $-\delta_{J_z}$.
Analogously, the phase shift in the
resonant-level model (\ref{RL}) is $\delta_W$ if the $d$-level is
unoccupied and $-\delta_W$ if the $d$-level is occupied.
For a clear distinction, here and in the following we denote
by $\rho_F$ a density of states of a fermionic band with finite cutoff,
whereas $\rho$ refers to a density of states within the cutoff scheme
underlying bosonization.
In the latter scheme, the phase shifts defined above
are directly proportional to the coupling constants, $\delta_{J_z}=
\pi J_z \rho/2$ and $\delta_W= \pi W \rho/2$, where the density of
states $\rho = 1/(2\pi v_{\rm F})$.
If one uses instead a model where
the high-energy cutoffs arise from a band-structure, one obtains
$\delta_{J_z}=\arctan \frac{(-J_z/2) \text{Im} g_{00}(0)}{1-(-J_z/2)
  \text{Re} g_{00}(0)}$, where $g_{00}(\w)=\sum_k
\frac{1}{\w-\epsilon_k+i 0^+}$ is the local Green's function of the
electrons.
In case of particle--hole symmetric bands,
$g_{00}(0) = -i \pi \rho_F$, this relation simplifies to
\begin{equation}
\delta_{J_z}=
\arctan[ \pi J_z \rho_F/2 ].\label{phaseJ}
\end{equation}
Similarly, $W$ in (\ref{Anderson}) induces for $V=0$ a phase shift $\delta_{W}=
\arctan[ \pi W \rho_F/2 ]$.
Matching the various models by their phase shifts, the relation
$W \rho= \sqrt{2}J_z \rho - (\sqrt{2} - 1)$ derived within bosonization, translates into
\begin{equation}\label{matchingJW}
\delta_{W}=
\arctan[ \pi W \rho_F/2 ]=\delta_{J_z}\sqrt{2} -\frac{\pi}{2} (\sqrt{2} -
1).
\end{equation}
Note that this equation is only valid for small $J_\perp$ and $V$.


\section{Strong-Coupling Analysis}
\label{sec:StrongCoupling}

In this section we analyze the behavior of the
system for small $T_K^{(1)} / K_z$.
After presenting a general argument for the
existence of a phase transition, we
discuss the resulting physics in terms of
the generalized Anderson model (\ref{Anderson}).
Interestingly, two {\em different} strong-coupling
limits emerge which will be described in
Secs.~\ref{sec:largeK} and \ref{sec:largeW}.
As detailed below, both strong-coupling limits display
a phase transition of the Kosterlitz--Thouless type.
Furthermore, the limits will be shown to commute,
and the physical regimes are smoothly connected.

\subsection{Effective Hamiltonian}
\label{sec:hstrong}

To investigate the phase diagram sketched in
Fig.~\ref{fig:phd1} we consider first the limit of
$|K_z|\to\infty$.
We can restrict the considerations to $K_z\to +\infty$,
as results for $K_z\to-\infty$ are similar because the $z$-component of
the total spin is conserved separately for the ``1'' and ``2''
subsystems.

In the limit $K_z\to +\infty$ the two
impurity spins form an antiferromagnetic mini-domain, with configurations
$|\!\uparrow\downarrow\rangle$ and $|\!\downarrow\uparrow\rangle$.
No fluctuations can occur for $K_z=\infty$ (or $J_\perp = 0$),
therefore the ground state of the full system is a doublet.

We now set up a perturbation theory in the small parameter $J_\perp / K_z$,
by deriving an effective Hamiltonian in the
\{$|\!\uparrow\downarrow\rangle$, $|\!\downarrow\uparrow\rangle$\}
subspace of the impurities.
The lowest process connecting
the two states $|\!\uparrow\downarrow\rangle$, $|\!\downarrow\uparrow\rangle$
is of order $\mathcal{O}(J^2_\perp/K_z)$.
Thus the effective Hamiltonian in the strong Ising-coupling limit
reads
\bea \label{StrongCouplingHamiltonian}
H^{\rm K}_{\rm eff} = H^{\rm K}_{\rm eff,0} + H^{\rm flip}_{\rm eff}
\eea
where $H^{\rm K}_{\rm eff,0}$ is given by $H_1^K+H_2^K$ with the
perpendicular Kondo coupling set to zero, $J_\perp =0$.  Note that the size of $J_z$ can be arbitrary.
The mini-domain is flipped by the term
\bea \label{DomainFlip}
H^{\rm flip}_{\rm eff} &=&
\frac{4 J^2_\perp}{K_z}
\left(
S_1^+ S_2^-
  \Psi^\dag_{\downarrow 1} \Psi^\pdag_{\uparrow 1}
  \Psi^\dag_{\uparrow 2} \Psi^\pdag_{\downarrow 2}
 + {\rm
    h.c.}
\right)
\eea
with $\Psi_{\alpha i} = \sum_{k} c^\pdag_{k \alpha i}$.
The zero-temperature stability of the frozen mini-domain
now depends on whether the operator $H^{\rm flip}_{\rm eff}$
is relevant in the renormalization group sense.

Since $H^{\rm flip}_{\rm eff}$ is comprised of four electron operators,
its bare (tree-level) scaling dimension is negative,
${\rm dim} [ H^{\rm flip}_{\rm eff} ]_{\rm tree} = -1$.
This might suggest that the doublet ground state with residual
entropy $\ln 2$ is stable.
However, in the present problem $H^{\rm flip}_{\rm eff}$ acquires
an anomalous scaling dimension which modifies this conclusion.
This can be understood as follows:
For large $J_z$ a flip of the mini-domain suddenly changes the phase
shifts of all electrons in the leads, thus exciting an infinite number
of particle--hole pairs.
This is the well-known orthogonality catastrophy,\cite{Anderson}
leading to an anomalous long-time response of the electrons.
In the presence of a sharp Fermi edge this results in a
so-called x-ray edge singularity which is reflected in an
anomalous scaling dimension of
$H^{\rm flip}_{\rm eff}$ (\ref{DomainFlip}).

In the following we will determine this scaling dimension using
Hopfield's rule of thumb,\cite{Hopfield} and identify the critical
$J_z$ where the $H^{\rm flip}_{\rm eff}$ becomes relevant,
resulting in ``quantum-melting'' of the frozen mini-domain.

To adjust the Fermi sea
to a new ground state after the domain has flipped once, a certain
amount of charge $\Delta n$ has to flow to infinity.  Hopfield noticed
that collective response of a Fermi sea depends in the long-time limit
only on this $\Delta n$: the corresponding correlation function decays
as $t^{-(\Delta n)^2}$.
In our problem, we have to consider four different Fermi surfaces
($j=1,2$, $\sigma=\uparrow,\downarrow$) each contributing independently.

A domain flip is induced by the operator
$A = S_1^+ S_2^- \Psi^\dag_{\downarrow 1} \Psi^\pdag_{\uparrow 1}
\Psi^\dag_{\uparrow 2} \Psi^\pdag_{\downarrow 2}$. According to Hopfield's rule of thumb,
the correlation function in the {\em absence} of domain flips is then given by
\bea \label{xray}
\langle A^\dag(t) A(0) \rangle_{H^{\rm K}_{\rm eff,0}} \sim t^{-\alpha} \,,
\eea
where \bea \alpha = \sum_{j=1,2;\sigma=\uparrow,\downarrow} \Delta
n^2_{j\sigma}\,.  \eea
The transferred charges $\Delta n_{j\sigma}$ are
easily obtained from the phase shifts using Friedel's sum rule. For
example, a spin flip induced by $A$ changes the phase of the down
electrons in bath $2$ from $\delta_{J_z}$ to $-\delta_{J_z}$ which
corresponds according to Friedel's sum rule to a charge transfer of $2
\delta_{J_z}/\pi$. Furthermore, the annihilation operator
$\Psi^\pdag_{\downarrow 2}$ eliminates one charge and the total charge
transfer in this channel is given by $\Delta n_{\downarrow 2} = 2
\delta_{J_z}/\pi -1$.  Similar arguments give $\Delta n_{\uparrow 2} =
-\Delta n_{\uparrow 1} = \Delta n_{\downarrow 1} = -\Delta n_{\downarrow
  2}$.  Therefore, the exponent $\alpha$ is given by $\alpha = 4
\left(\frac{2 \delta_{J_z}}{\pi} -1 \right)^2$.
This result can also be verified explicitly by bosonization
following Schotte and Schotte.\cite{Schotte^2}

From Eq. (\ref{xray}),  we can directly read off
the anomalous scaling dimension of
the domain-flip Hamiltonian (\ref{DomainFlip}) with
respect to the ``frozen-domain'' fixed point:
\bea \label{dimHflip}
{\rm dim} [H^{\rm flip}_{\rm eff}] =
1 - \frac{\alpha}{2} = 1 - 2 \left(\frac{2 \delta_{J_z}}{\pi} -1 \right)^2\,.
\eea
Here, the first term arises from the engineering dimension of $H^{\rm flip}_{\rm eff}$.
For small scattering phase shifts ${\rm dim} [H^{\rm flip}_{\rm eff}]$  is negative,
i.e., the domain flip Hamiltonian is
irrelevant and the doubly degenerate ``frozen-domain'' fixed point described by
$H^{\rm K}_{\rm eff,0}$ is stable.
Domain flips become relevant for $\delta_{J_z}>\delta_T$ with
\bea \label{deltaTdef}
\delta_T = \frac{\pi}{2} \left(1 - \frac{1}{\sqrt{2}}\right)\,.
\eea
Beyond $\delta_T$ the fluctuations of the mini-domain grow towards low
energies giving rise to a new phase -- in this regime the
pseudospin describing the mini-domain is still well defined,
but is ultimately screened at low energies.
The special value of $\delta_T$ is well-known as the Toulouse point
of the single-impurity Kondo Hamiltonian.\cite{GogoBook}

We have thus established the existence of a phase transition
in the limit of $J_\perp/K_z \ll 1$ which is accessed by varying
$J_z$.
In the bosonization cutoff scheme, the critical value is
$\rho J_z^{cr} = 1 - 1/\sqrt{2}$ corresponding to the Toulouse
point of the individual Kondo impurities in the original
model (\ref{Model}).

\subsection{Relation to the generalized single-impurity model}

What are the properties of this
``fluctuating mini-domain'' and what is the nature of the
quantum phase transition separating the two phases?
This question can be tackled within the generalized Anderson
model (\ref{Anderson}).

According to Eq.~(\ref{matchingJW}) the coupling $W$
in (\ref{Anderson}) vanishes for $\delta_{J_z}=\delta_T$, i.e.,
at the Toulouse point.
This does, however, not necessarily imply that $W$ can be treated as
a small parameter in the vicinity of the Toulouse point.
Rather, it is important to realize that two {\em different}
strong-coupling limits emerge, depending on the order of
limits taken in parameter space:

(i) If we consider the limit $W\to 0$ at fixed small $J_\perp$,
then the critical $K_z$ will diverge according to the analysis
above.
In this case, $K_z$ is the largest scale, whereas $J_\perp$ and $V$
can be treated as perturbations.

(ii) Alternatively, we can take $J_\perp\to 0$ at fixed small $W < 0$.
Apparently, the critical $K_z$ will {\em vanish}
(although $K_z / T_K^{(1)} \gg 1$), and a strong-coupling
treatment has to consider that $W$ is a large scale.

Both limits correspond to physical situations.
However, the scaling limit,
discussed at the beginning of Sec.~\ref{sec:Model}
and employed in plotting the phase diagram of Fig.~\ref{fig:phd1},
is reached by taking $J_\perp\to 0$ at fixed $J_z$ and
$T_K^{(1)} / K_z$, corresponding to the case (ii) above.
More generally,
in a three-dimensional phase diagram which involves a third
axis labelled $J_\perp$ in addition to the ones shown in
Fig.~\ref{fig:phd1}, case (i) applies for any finite $J_\perp$ in a
certain vicinity of the Toulouse point, before the behavior crosses
over to case (ii) -- this crossover scale vanishes in the universal
limit $J_\perp\to 0$.

We will now analyse both cases separately --
interestingly the two limits will turn out to commute.

\subsection{$W\to 0$ at fixed $V$}
\label{sec:largeK}

\begin{figure}
\centerline{
\includegraphics[width=3in]{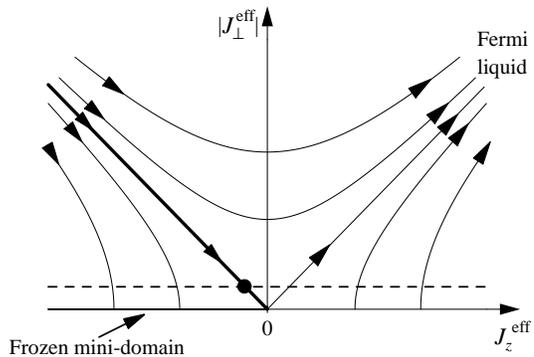}
}
\caption{
Schematic renormalization-group flow of the effective
Kondo model (\protect\ref{effKondo}),\cite{Hewson,RMP}
describing the screening of the mini-domain pseudospin in the
limit of large $K_z$.
Here, $J^{\text{eff}}_{z} = 0$ corresponds to the Toulouse
point of the original Kondo model.
The Fermi-liquid fixed point is characterized by pseudospin
screening; the frozen-mini-domain phase corresponds to
a {\em line of fixed points} with unscreened pseudospin.
The thick line denotes the phase boundary, given by
$J^{\text{eff}}_{z} = -|J^{\text{eff}}_\perp|$.
Variation of $J_z$ around the Toulouse point in the original
model (at fixed $J_\perp$ and $K_z$) corresponds to a parameter
variation in the effective model as shown by the dashed line;
the dot is the transition point.
The transition is in the Kosterlitz--Thouless universality
class.
}
\label{fig:kondoflow}
\end{figure}

Assuming that $K_z$ is the largest scale in the problem,
which corresponds to a strong local Coulomb repulsion on the impurity
site of the generalized Anderson model,
we can directly map it onto an anisotropic Kondo model.
With
$\bar{n}_{\Psi \sigma}=:\Psi^\dag_{\sigma}(0) \Psi^\pdag_{\sigma}(0)
:$ and $\vec{S}= \frac{1}{2} d^\dag_{\alpha}
{\boldsymbol{\sigma}}_{\alpha \beta} d^\pdag_{\beta}$, we can rewrite
$W \sum_\sigma \bar{n}_{d \sigma} \bar{n}_{\Psi \sigma}=W S_z
(\bar{n}_{\Psi \uparrow}-\bar{n}_{\Psi \downarrow})+\frac{W}{2}
(\bar{n}_{d \uparrow}+\bar{n}_{d \downarrow}) (\bar{n}_{\Psi
  \uparrow}+\bar{n}_{\Psi \downarrow})$. As charge fluctuations are
frozen out for large $K_z$, the last term can be omitted, $\bar{n}_{d
  \uparrow}+\bar{n}_{d \downarrow}=1$, and we obtain
\begin{equation}
\label{effKondo}
H^{A}_{\rm eff} = H_0[\Psi_{\sigma}] +
\sum_{n\alpha\beta} J^{\text{eff}}_{n} S^n
\Psi^\dag_{\alpha}(0) \sigma^n_{\alpha\beta} \Psi^\pdag_{\beta}(0) ,
\end{equation}
with $J^{\text{eff}}_{n}= (J^{\text{eff}}_\perp,J^{\text{eff}}_\perp, J^{\text{eff}}_z)$
where
\begin{equation}
J^{\rm eff}_\perp
= \frac{4 V^2}{K_z},\qquad J^{\rm eff}_z = \frac{4 V^2}{K_z}+ W.
\end{equation}
Spin-up and spin-down in this Hamiltonian correspond to the two states
of the mini-domain.
The phase diagram of (\ref{effKondo}) is well
known:
when $J_z^{\text{eff}}$ is increased from negative values towards zero
(keeping $J_\perp^{\text{eff}}\neq 0$ fixed)
one observes a quantum phase transition from a ferromagnetic
regime with $\ln 2$ residual entropy to a Fermi-liquid phase where the
spin is quenched, see Fig.~\ref{fig:kondoflow}.
The $S_0=\ln 2$ phase can obviously be identified with
our ``frozen mini-domain''. It is stable for
$J^{\text{eff}}_{z}<-|J^{\text{eff}}_\perp|$ or $W<W_c$ with
\begin{equation}
W_c=-\frac{8 V^2}{K_z}\,.  \label{wc}
\end{equation}
For $K_z \to \infty$ or $V\to0$, the phase transition is located at $W_c=0$ or equivalently
$\delta_{J_z}=\delta_T$ as anticipated above.
Eq. (\ref{wc}) is the {\em exact} result for the phase boundary
for $K_z\to\infty$, provided that $V$ and $W$ are mapped onto
$J_\perp$ and $J_z$ as described in Sec.~\ref{sec:Model}.
We note again that the limit considered here
does {\em not} correspond to the universal limit $J_\perp\to 0$,
because this would give $K_z^{cr}\to 0$ at any fixed $W$.

\subsection{$V\to 0$ at fixed $W$}
\label{sec:largeW}

Anticipating that $K_z^{cr}\to 0$ in this limit, we need
to consider a problem where $W$ is a large local energy scale.
Interestingly, at $K_z=0$ and $V=0$ the four impurity
states are degenerate even in the presence of $W$ because
of overall particle--hole symmetry. This degeneracy is
lifted by $K_z$ which (as above) favors the impurity states
$|\!\uparrow\downarrow\rangle$ and $|\!\downarrow\uparrow\rangle$
(assuming $K_z>0$).

We proceed with an analysis similar to the usual Schrieffer--Wolff
transformation, now in the presence of a large $W$.
As usual, hopping processes of second order in $V$ produce
an effective pseudospin interaction between the impurity
and the conduction band.
However, as the impurity occupation in the intermediate states
is different from that of initial and final states, the
intermediate state physics involves an x-ray edge singularity
due to the sign change of the strong potential scatterer $W$.

Analyzing the matrix elements, we find x-ray edge behavior which
is cutoff by $K_z$, i.e., the effective generated Kondo couplings
are of the form $V^2/K_z^{1-\beta}$ where
\begin{equation}
\beta = - 2 \frac{2}{\pi}\delta_W - \left(\frac{2}{\pi}\delta_W\right)^2
\end{equation}
where $\delta_W = \pi W \rho / 2$ is the phase shift from $W$.

The detailed derivation of the generalized Schrieffer--Wollf transformation is
given in App.~\ref{sect:SW}.
Finally, one arrives at an effective Kondo model of the form (\ref{effKondo}),
but now with couplings
\begin{equation}
J^{\rm eff}_\perp
= \frac{4 V^2\,f_\perp(\delta_W)}{K_z^{1-\beta} \Lambda^\beta} \,,\qquad
J^{\rm eff}_z = \frac{4 V^2 \, f_z(\delta_W)}{K_z^{1-\beta} \Lambda^\beta} + W
\end{equation}
where $\Lambda$ is of the order of the band cutoff, and
both $f_\perp$ and $f_z$ are smooth, dimensionless functions of $\delta_W$
with
\beq
\lim_{W\to 0} f_\perp(\delta_W) = f_\perp(0) = 1 \,,~
\lim_{W\to 0} f_z(\delta_W) = f_z(0) = 1 \,.
\label{flimit}
\eeq
As above, the model shows a Kosterlitz--Thouless phase transition
in this effective Kondo model occurs at the
line $J^{\rm eff}_\perp = -|J^{\rm eff}_z|$, i.e.
\begin{equation}
\label{wimpl}
W = - \frac{4V^2 [f_\perp(\delta_W) + f_z(\delta_W)]}{K_z^{1-\beta} \Lambda^\beta}
\end{equation}
where $\beta$ is a function of $W$ according to $1-\beta = 2
(1\!-\!\alpha)$ with $\alpha = (J_z \rho - 1)^2$ as introduced
above.
To obtain an explicit relation between $W$, $V$, and $K_z$ in
the vicinity of the Toulouse point, we expand in $W$.  Dropping
additive logarithmic corrections, we have
\begin{equation}
W^\frac{1}{1-\beta} \approx W \,, ~~
\Lambda^{\beta} \approx 1 \,.
\end{equation}
With this and (\ref{flimit}) we can re-write Eq. (\ref{wimpl}) as
\begin{equation}
\label{wc2}
W
= - 8 \frac{V^\frac{1}{1-\alpha} \Lambda^\frac{1-2\alpha}{1-\alpha}}{K_z}
\end{equation}
which is {\em smoothly} connected to the result (\ref{wc})
obtained in the limit $W\to 0$ of Sec.~\ref{sec:largeK}.

As announced, the critical $K_z$ for fixed $W$ depends only on
$T_K^{(1)} \propto V^\frac{1}{1-\alpha}
\Lambda^\frac{1-2\alpha}{1-\alpha}$. To fix the prefactor, we first
have to {\em define} $T_K^{(1)}$ for an asymmetric Kondo model.
This is best done using a physical observable like the impurity
specific heat coefficient $\gamma=\lim_{T\to 0} C_{\rm imp}/T$ of the
anisotropic single-impurity Kondo model.
We employ
\begin{equation}
\label{tkdef}
T_K^{(1)} \equiv w \frac{\pi^2}{3} \gamma^{-1}\,,
\end{equation}
where $w=0.4128$ is the Wilson number.\cite{Hewson}
At the Toulouse point, $\alpha = 0$,
one easily finds $\gamma=1/(3 \rho_F V^2)$.
Thus, for $W \to 0$ we obtain:
\begin{equation}
\label{wc3}
\rho_F W
= - \frac{8}{ w \pi^2} \frac{T_K^{(1)}}{K_z^{cr}}\approx
-1.964  \frac{T_K^{(1)}}{K_z^{cr}}.
\end{equation}
For particle--hole symmetric bands, this can be re-written using Eqs. (\ref{phaseJ}) and
(\ref{matchingJW}) into:
\begin{eqnarray}
\frac{T_K^{(1)}}{K_z^{cr}}&=&\frac{\rho_F  w \pi^2 \sqrt{2} \sin^2(\frac{\pi}{2 \sqrt{2}})}{8} \left(J_z^{cr}-J_z \right) \nonumber \\
&\approx& 0.578 \, \rho_F \left(J_z^{cr}-J_z \right)      \label{wc3J}
\end{eqnarray}
valid for $J_z \to J_z^{cr}$.

Concluding this analysis, we have shown that also in the
limit $V\to 0$ (keeping $W$ fixed)
the effective Anderson model (\ref{Anderson}) can be mapped onto
an effective Kondo model, which describes the screening of the
mini-domain pseudospin. The phase transition is in
the Kosterlitz--Thouless universality class.
The two strong-coupling limits are adiabatically connected, as
the involved impurity states and transitions are similar in both
cases; in addition the equations for the phase boundaries
(\ref{wc}) and (\ref{wc2}) match.

From the mapping to the anisotropic Kondo model we can also identify
the ``fluctuating mini-domain'' regime with a Fermi-liquid phase.
Here, the pseudospin is screened below the Kondo temperature,
$T^\ast$, of the effective Kondo model (\ref{effKondo}).  Importantly,
the Fermi sea of the effective model is formed by solitonic spin
excitations of the original model (\ref{Model}) -- this will strongly
influence the conductance through the system as discussed in
Sec.~\ref{sec:Appl}.  At the Toulouse point, we can estimate $T^\ast
\sim {\rm min}(K_z,D) \exp[-|K_z|/(8V^2\rho_F)]$.
Close to the quantum phase transition,
reached for $K_z/V^2 \to 0$ at the Toulouse point,
$T^*$ is exponentially small.
Similiarly, for finite $K_z^{cr}$ and $K_z\lesssim K_z^{cr}$,
one expects for a Kosterlitz--Thouless transition the behavior \cite{Koster74}
\begin{equation}
T^* = a e^{-b/\sqrt{K_z^{cr}-K_z}},
\label{tstarkt}
\end{equation}
where $a$ is a function of $T_K^{(1)}$ and $J_z$;
this form will actually be used to fit the numerical data
of Sec.~\ref{sec:NRG}.

At the Toulouse line, one can investigate the crossover from the
``fluctuating mini-domain'' regime to the regime with Kondo-screened
spins at $K_z=0$ (dashed crossover line in Fig.~\ref{fig:phd1}).
As our model is equivalent to an Anderson impurity model (\ref{Anderson}),
this crossover is equivalent to the well-known Anderson model crossover
from mixed-valence to Kondo behavior.
This crossover takes place at $K_z \sim V^2 \rho_F$, i.e., when
$T_K^{(1)}/K_z$ is of order ${\cal O}(1)$. Similiarly, one finds for $J_z\to
\infty$ that this crossover is located at $T_K^{(1)}\sim J_\perp
\sim K_z$, and furthermore for $J_z\to 0$ one also expects this
crossover at $K_z \sim T_K^{(1)}$
(as no other low-energy scale exists in this regime).
We therefore conclude that the dashed crossover line in
Fig.~\ref{fig:phd1} is always located at $T_K^{(1)}/K_z\sim {\cal O}(1)$.

The schematic RG flow of the effective Kondo model, shown in
Fig.~\ref{fig:kondoflow}, illustrates the behavior of the
Ising-coupled two-impurity model in the strong-coupling regime.  Three
observations are important: (i) Both the $S_0 =\ln 2$ frozen-domain
phase and the $S_0=0$ Fermi-liquid phase are stable to small
perturbations -- this conclusion is in agreement with the analysis of
Ref.~\onlinecite{Andrei}.
(ii) The $S_0=\ln 2$ phase corresponds to a {\em line} of RG fixed
points.
(iii) The two phases are separated by
a Kosterlitz--Thouless transition, characterized by logarithmic
rather than power-law behavior of thermodynamic observables.

So far, the conclusions above mainly apply to the vicinity of the Toulouse
point of the single-impurity model, i.e., for strongly anisotropic Kondo
coupling.
In the next section we will present numerical results which strongly
support that the above picture is valid over the whole phase diagram.
In particular, we shall show that the phase transition is of Kosterlitz--Thouless
type even in the case of isotropic Kondo couplings.
Furthermore, no other phase transition (than the one indicated in Fig.~\ref{fig:phd1})
occurs; this implies that the Fermi liquid formed by the fluctuating
mini-domain is adiabatically connected to the Fermi liquid of two
individually Kondo-screened spins, realized in the limit
$K_z \ll T_K^{(1)}$.


\section{Numerical Renormalization Group analysis}
\label{sec:NRG}

In this section, we turn to a numerical investigation of
the model of Ising-coupled impurities using the
numerical renormalization group (NRG) technique.\cite{Wilson75}
In principle, an investigation of the original
two-impurity model (\ref{Model}) is possible -- however,
the required two bands of spinful fermions are computationally demanding
within NRG, and results are significantly less accurate compared to
the NRG treatment of single-band impurity models.

Therefore, we have decided to study the generalized
Anderson model of Eq. (\ref{Anderson}), obtained after
bosonization and refermionization of the original model.
Featuring only one band of spinful conduction electrons,
it allows high-accuracy numerical simulations down to lowest
energy scales and temperatures.

\subsection{Parameters}

In the following, we will show numerical results for
the generalized Anderson model (\ref{Anderson}), for
a rectangular particle--hole symmetric fermionic band of width
$D=2\sqrt{2}$, and different values of the hybrization $V$,
density interaction $W$, and on-site repulsion $K_z$ --
note that $K_z$ is identical to the original Ising interaction
between the two impurities, whereas $V$ and $W$ are related to
$J_\perp$ and $J_z$ as detailed in Sec.~\ref{sec:Model}.
In particular, $W$ measures the deviation from the Toulouse
point.
According to Eq.~(\ref{matchingJW}), valid for small $V$,
$J_z = 0$ corresponds to
$W\rho_F = - (2/\pi)\tan[\pi(\sqrt{2}-1)/2] \approx -0.48$.
Similarly, large negative values of $W$, i.e., $W\to -\infty$,
correspond to
$\rho J_z = 1-\sqrt{2} = - \sqrt{2} \,(\rho J_z)_{\rm Toulouse}$
with $\rho J_z$ here defined in the bosonization cutoff scheme.

The mapping between $V$ and $J_\perp$ (which are simply
proportional) can be achieved via the Kosterlitz--Thouless
transition line for a single impurity:
This line, where $T_K^{(1)}$ vanishes, is given by $J_z = -|J_\perp|$.
Using NRG, we have numerically determined a few points on this line,
characterized by parameters $V$ and $W$ in the model formulation
(\ref{Anderson}) with $K_z = 0$ -- note that this involves
an extrapolation of $T_K(V)$ to $T_K=0$.
With the mapping between $W$ and $\rho J_z$ established above,
we find for the employed parameters the correspondence
$\rho J_\perp \approx 0.38 V$, valid for small $V$.

Within NRG, the bath density of states is discretized on a logarithmic
grid, with a discretization parameter $\Lambda$, i.e.,
the energy axis is divided into intervals at points
$D, D\Lambda^{-1},D\Lambda^{-2}, \ldots$.
The discretized model is transformed into a semi-infinite chain form,
and then diagonalized iteratively.\cite{Wilson75}
After each diagonalization step the lowest $N_s$ eigenstates of the
Hamiltonian are kept.
Clearly, the results are ``exact'' in the limit $N_s\to\infty$, $\Lambda\to 1$.
In practice, numerical results are monitored at fixed $\Lambda$, where
convergence upon increasing the value of $N_s$ can be readily achieved.
Then, these converged numbers are extrapolated as function of
$\Lambda$ to $\Lambda=1$ to obtain an estimate for the ``exact''
result.

Most of the NRG calculations here have employed a
discretization parameter of $\Lambda\!=\!2$,
keeping $N_s\!=\!650$ states per NRG step.
For selected parameter values we have performed calculations with
different $\Lambda$ and $N_s$.
For each $\Lambda$ convergence
w.r.t. $N_s$ can be easily achieved, however, for some
quantities the $\Lambda$ dependence turns out to be rather strong.
Therefore, the $\Lambda\to 1$ extrapolation has to be performed carefully,
as will be detailed at the end of this Section.

The results shown below primarily correspond to the universal regime
of $V\ll W,D$ and $K_z \ll W,D$; we have also performed some calculations
in the regime $K_z \gg W$ (not shown) with results consistent with the analysis
of Sec.~\ref{sec:largeK}.

\subsection{Results for RG flow and entropy}

\begin{figure}
\centerline{
\includegraphics[width=3.3in]{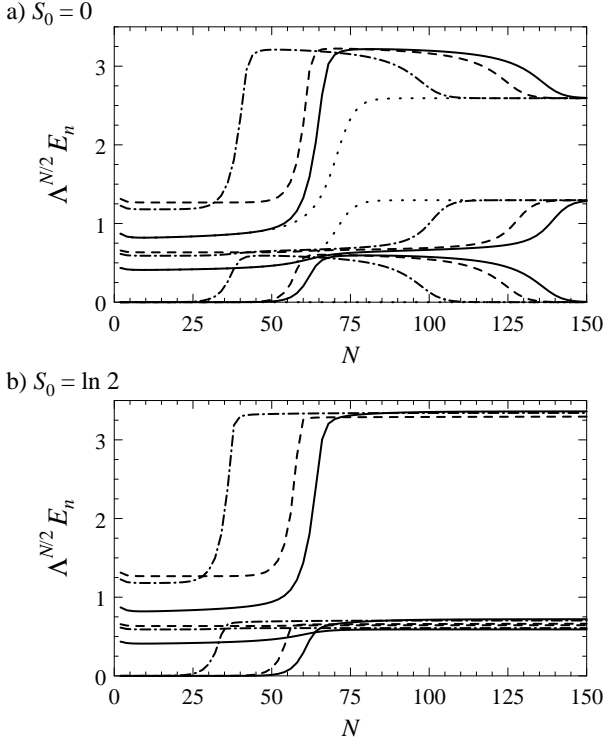}
}
\caption{
NRG flow diagram for the generalized single-impurity
Anderson model (\protect\ref{Anderson}),
for parameter values belonging to
a) the Fermi-liquid phase with $S_0=0$,
b) the frozen-domain phase with $S_0=\ln 2$.
Solid:    $W\rho_F = -0.44$, $V = 0.075$               ($K_z^{cr} = 7.6 \times 10^{-10}$),
Dash-dot: $W\rho_F = -0.10$, $V = 1.5 \times 10^{-3}$  ($K_z^{cr} = 3.4 \times 10^{-6}$),
Dashed:   $W\rho_F = -0.034$, $V = 1.5 \times 10^{-5}$ ($K_z^{cr} = 4.8 \times 10^{-9}$).
In a) and b), $K_z$ has been chosen slightly below and above
the critical value, respectively.
For all parameters, the system is in a $S=\ln 4$ regime
at high temperatures (small $N$), in a) it flows to
the $S=0$ state by passing through a regime with $S=\ln 2$.
In a), the additional dotted curves show the flow
for $W\rho_F = -0.44$, $V = 0.075$, and $K_z=0$.
The $W\rho_F$ values span a large range of anisotropies;
nevertheless, the $S=0$ fixed point is unique, and
the finite-temperature crossover is universal
for the curves close to $K_z^{cr}$.
Panel b) nicely shows that $S=\ln 2$ actually corresponds
to a line of fixed points.
NRG parameters are $\Lambda\!=\!2$ and $N_s\!=\!650$.
}
\label{fig:lvl}
\end{figure}

In Fig.~\ref{fig:lvl} we show NRG flow diagrams displaying the energies
of a few low-lying many-body eigenstates as function of
the number of NRG steps.
The data in Fig.~\ref{fig:lvl}a clearly show
that for small values of $K_z$ the same fixed
point is reached for any $V$ and $W$ -- this fixed point can be identified
with the Fermi-liquid phase, which is in particular also reached for $K_z=0$.
Therefore, the Fermi-liquid regime of two separately Kondo-screened impurities
is adiabatically connected to the ``fluctuating mini-domain'' regime
which can be characterized by pseudospin screening below the collective
Kondo temperature $T^\ast$.
In Fig.~\ref{fig:lvl}b flow diagrams for larger values of $K_z$ are shown --
here the fixed points reached at low energies are very similar for different
parameter sets, but not identical -- this is consistent with the notion
of a line of fixed points with $\ln 2$ residual entropy.

\begin{figure}
\centerline{
\includegraphics[width=3.3in]{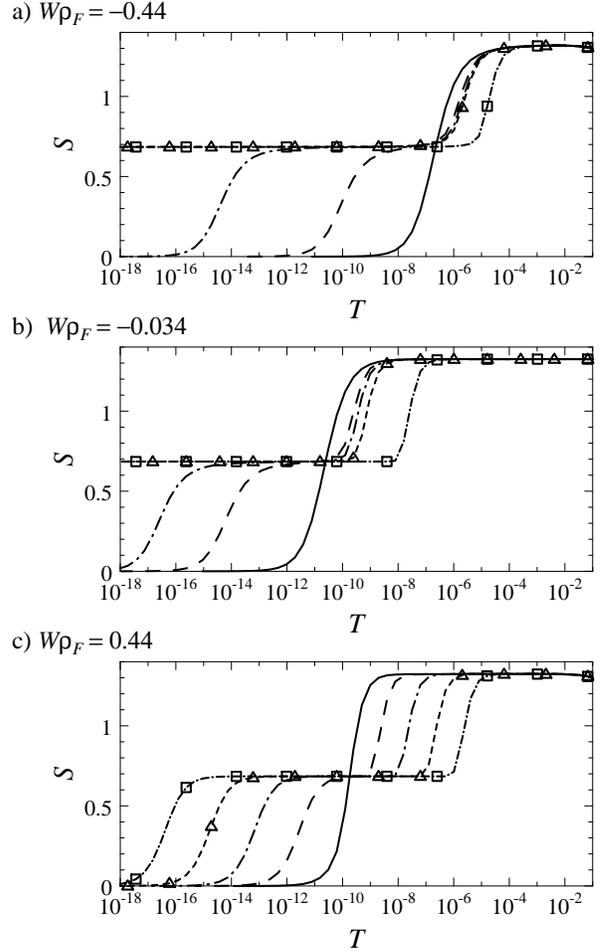}
}
\caption{
  Temperature evolution of the impurity entropy calculated by NRG for
  the generalized single-impurity Anderson model
  (\protect\ref{Anderson}) for different anisotropies of
  the Kondo coupling.
  In the ``frozen mini-domain'' phase
  the residual entropy is $\ln 2$ while it vanishes for
  $K_z<K_z^{cr}$. For $T_K^{(1)}\gg K_z$ (solid curves),
  the high-temperature $\ln 4$ entropy is quenched in a single step,
  whereas two-stage screening occurs for $T_K^{(1)} < K_z < K_z^{cr}$.
a) $W\rho_F = -0.44$, $V = 0.15$               ($K_z^{cr} = 1.5 \times 10^{-5}$),
close to isotropic Kondo coupling.
$K_z$ is:
Solid 0, long-dash $10^{-5}$,
long-dash-dot $1.3 \times 10^{-5}$,
short-dash $1.5 \times 10^{-5}$,
short-dash-dot $10^{-4}$.
b) $W\rho_F = -0.034$, $V = 1.5 \times 10^{-5}$ ($K_z^{cr} = 4.8 \times 10^{-9}$),
i.e., close to the Toulouse point of the individual Kondo impurities.
The $K_z$ values are:
Solid 0, long-dash $10^{-9}$,
long-dash-dot $1.5 \times 10^{-9}$,
short-dash $3 \times 10^{-9}$,
short-dash-dot $10^{-7}$.
c) $W\rho_F = 0.44$, $V = 1.5 \times 10^{-7}$, i.e., on the
right-hand side of the phase diagram Fig.~\protect\ref{fig:phd1}
where no phase transition occurs as function of $K_z$.
$K_z$ is:
Solid 0, long-dash $10^{-8}$,
long-dash-dot $10^{-7}$,
short-dash $10^{-6}$,
short-dash-dot $10^{-5}$.
}
\label{fig:entr}
\end{figure}

It is important to emphasize that no additional fixed point is
observed for $K_z \approx K_z^{cr}$, which could possibly correspond
to a critical fixed point.
This clearly shows that the quantum phase transition in our problem
is not associated with standard critical behavior, but indicates that it is
of the Kosterlitz--Thouless universality class.

To characterize the fixed points, we have evaluated the impurity entropy $S(T)$
using NRG. In Fig.~\ref{fig:entr} we show results for different values
of $W$ and several $K_z$.
The discussed two-stage quenching of the entropy, occuring
for $T_K^{(1)} < K_z < K_z^{cr}$, can be nicely seen in all
panels -- note that panel c) shows data for $W>0$, i.e.,
on the right-hand side of the Toulouse point.

\subsection{Phase boundary and phase diagram}\label{sec:NRG:boundary}

\begin{figure}
\centerline{
\includegraphics[width=3.4in]{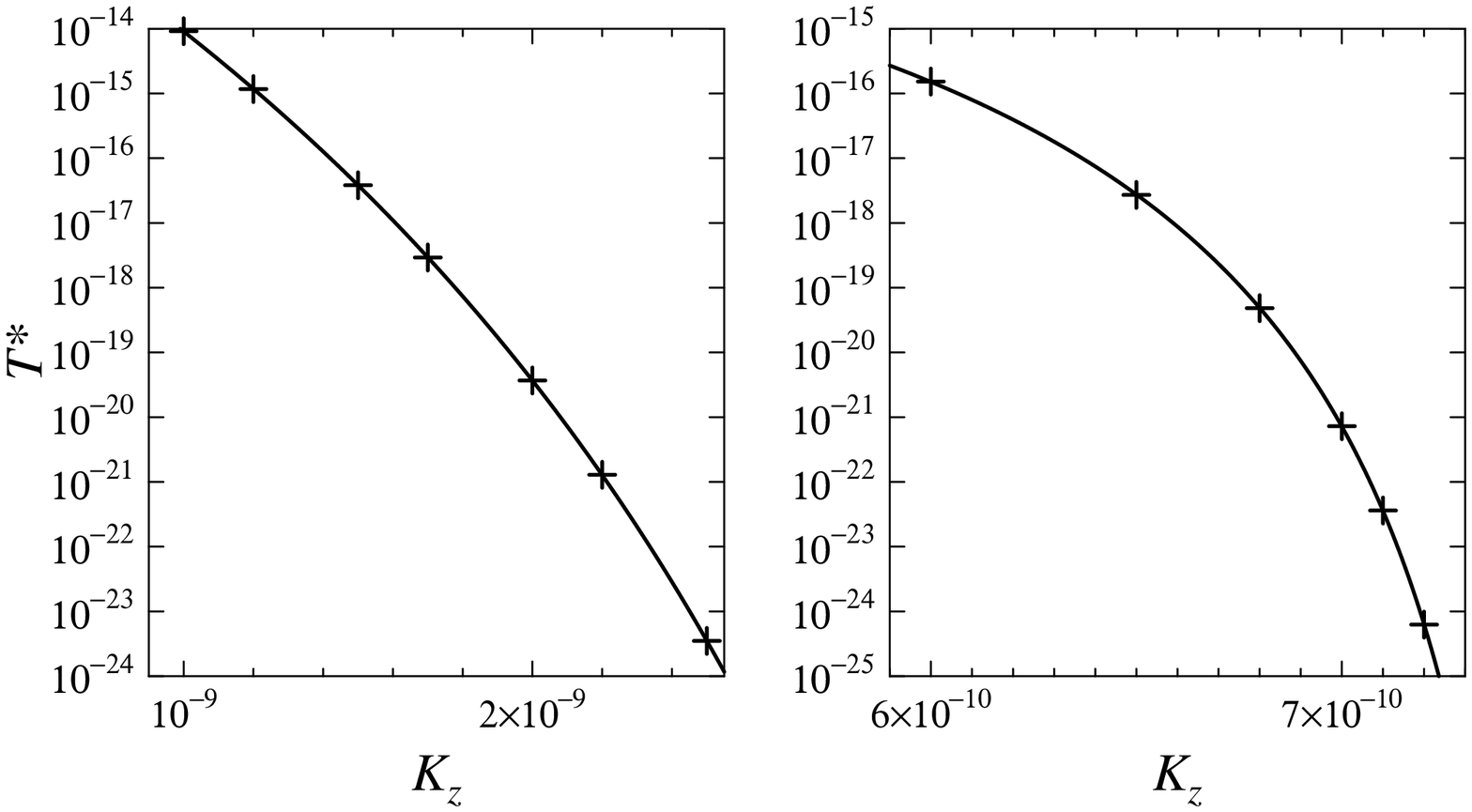}
}
\caption{
Numerically determined values of $T^\ast(K_z)$ for
$W\rho_F = -0.034$, $V = 1.5 \times 10^{-5}$ where $K_z^{cr} = 4.8 \times 10^{-9}$ (left),
and
$W\rho_F = -0.44$, $V = 0.075$ where $K_z^{cr} = 7.6 \times 10^{-10}$ (right);
together with the exponential fit described in the text.
}
\label{fig:kzfit}
\end{figure}

From the NRG results for fixed values of $W$, $V$
and different $K_z<K_z^{cr}$ it is possible to extract
a characteristic crossover temperature $T^\ast$
below which the pseudospin is screened, see above.
For numerical simplicity we defined $T^\ast$ through
$S(T^\ast) = 0.4$.
The dependence of $T^\ast$ on $K_z$ allows to determine
$K_z^{cr}$ where $T^\ast$ vanishes.
We fitted the data with $T^\ast(K_z) = a \exp[b /\sqrt{K_z^{cr} - K_z} ]$
(\ref{tstarkt})
with fit parameters $a$, $b$, $K_z^{cr}$,
which is the form expected near a Kosterlitz--Thouless transition \cite{Koster74}.
The fit works excellently for all anisotropies, as shown in
Fig.~\ref{fig:kzfit} -- this is again strong support for
the Kosterlitz--Thouless nature of the transition.

The single-impurity Kondo temperature, $T_K^{(1)}$ for given
$W$ and $V$ is determined from Eq.~(\ref{tkdef}), where
the specific heat coefficient $\gamma$ is extracted from
the NRG data for $S(T)$.

\begin{figure}
\centerline{
\includegraphics[width=3.2in]{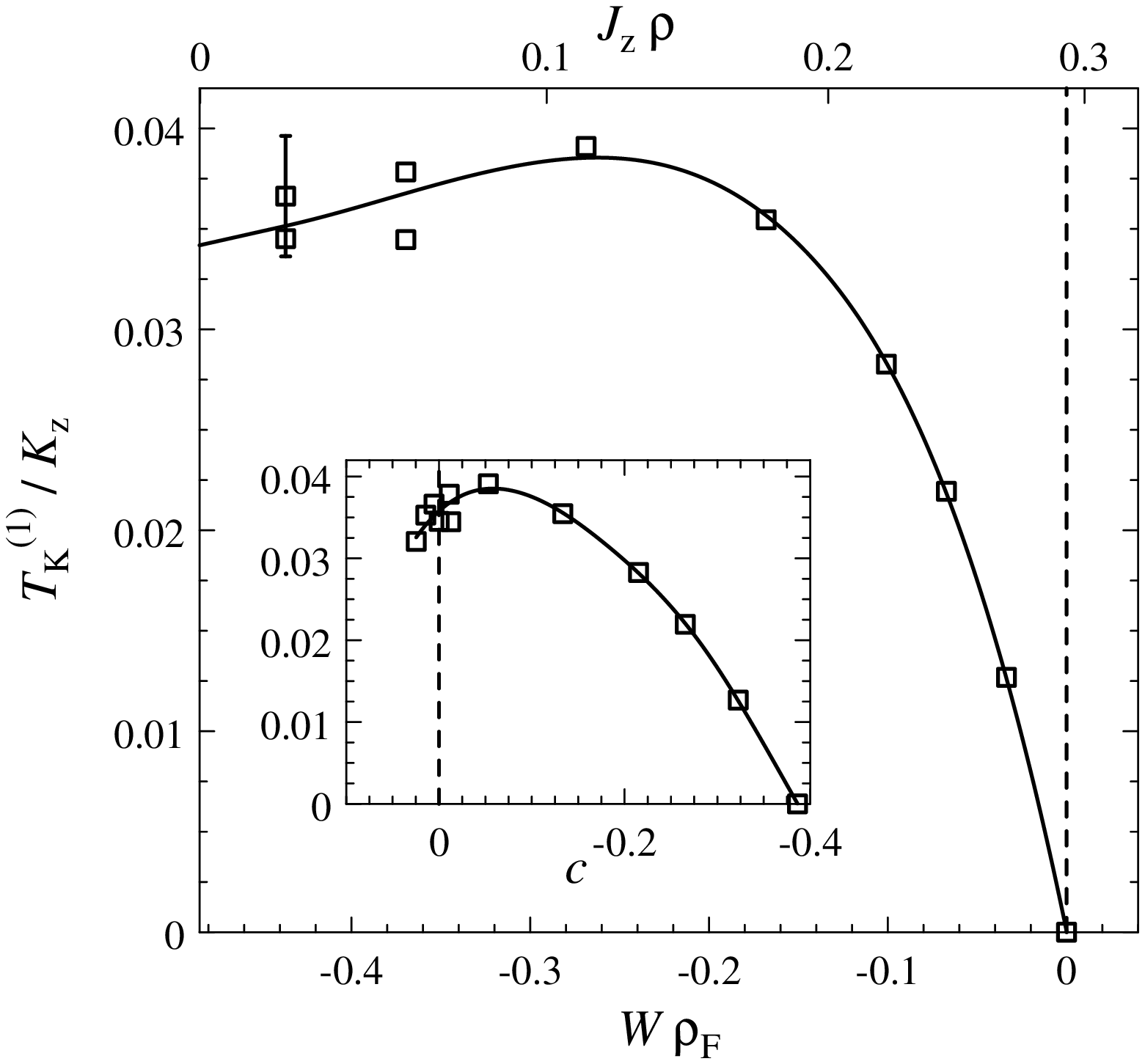}
}
\caption{
Phase diagram of the generalized single-impurity
Anderson model (\protect\ref{Anderson})
deduced from NRG calculations for NRG
discretization parameter $\Lambda=2$.
The vertical dashed line shows the Toulouse point of the
individual Kondo impurities.
Small values of $V$ have been used to reach the universal
regime $T_K^{(1)} \ll D$.
Precise values of $T_K^{(1)}$ have been determined via
the specific heat coefficient $\gamma$, see text.
The upper horizontal axis shows the corresponding values
of $J_z$ in the bosonization cutoff scheme.
The error bar shows the typical uncertainty in the
numerical determination of $T_K^{(1)}/K_z^{cr}$
arising from the fits of both $\gamma$ and $K_z^{cr}$.
The inset shows the same data for $T_K^{(1)}/K_z^{cr}$,
now plotted as function of the RG invariant $c$
of the single-impurity model (\ref{rginv}) --
this plot covers the range of positive as well as negative $J_z$
(here $c>0$).
The lines are guide to the eye only.
}
\label{fig:nrgphd1}
\end{figure}

Having determined both $T_K^{(1)}$ and $K_z^{cr}$, we are in the position
to plot the phase diagram in the universal fashion indicated
in Fig.~\ref{fig:phd1}, i.e., employing the limit $T_K^{(1)} \ll D$.
The result is shown in the main panel of Fig.~\ref{fig:nrgphd1}.

It is possible to make the meaning of universality more precise:
so far we have distinguished the parameter sets by
their value of $J_z$ (or $W$), leading to different values
of $T_K^{(1)}/K_z^{cr}$.
Within the RG treatment of Yuval and Anderson \cite{Yuval70}
for the single-impurity Kondo model,
it is easily seen that we can expect identical low-energy
behavior for two single-impurity models if the initial
parameters place the two models on the same RG trajectory.
(The RG trajectories are identical to the ones shown in
Fig.~\ref{fig:kondoflow}.)
Therefore, the correct parameter for the horizontal axis of
our phase diagram is a parameter labelling the RG trajectories,
i.e., a proper RG invariant.
Note that for $J_z>0$ and $J_\perp\to 0$, $J_z$ can be used
as such a label -- this is what we have done so far.
A proper RG invariant is $c$ defined by \cite{Yuval70}
\bea
\label{rginv}
c &=& 4 (J_\perp \rho)^2 + \epsilon + 2 \ln \left(1-\frac{\epsilon}{2}\right) \,, \\
\epsilon &=& 8 \frac{\delta_{J_z}}{\pi} - 8 \left(\frac{\delta_{J_z}}{\pi}\right)^2 \,, \nonumber
\eea
where $\delta_{J_z} = \pi J_z \rho / 2$ is the phase shift resulting from $J_z$,
and we have employed the bosonization cutoff scheme here.
For small values of both $J_z$ and $J_\perp$, the above equations
can be expanded to yield:
\begin{equation}
c = 4 (J_\perp \rho)^2 -  4 (J_z \rho)^2 \,.
\end{equation}
For $J_\perp\to 0$ the value of $c$ thus depends only on $J_z$ as
anticipated;
in this regime $c<0$.
The advantage of using a parametrization of the single-impurity
RG flow via $c$ is that it allows to cover the trajectories
with $|J_\perp| > |J_z|$ as well, i.e., the trajectories
above the isotropic line in Fig.~\ref{fig:kondoflow};
here $c>0$.

With the parameter mapping described at the beginning of this section
we have all parameters at hand and can determine the value of
$c$ from $V$ and $W$.
This allows to re-plot the phase diagram in a plane
spanned by $T_K^{(1)}/K_z^{cr}$ and $c$ -- this is
shown in the inset of Fig.~\ref{fig:nrgphd1}.
In particular, we can now add data points for $J_z < 0 $ to the
phase boundary plot, as those are characterized by a
{\em finite} $J_\perp$ in the limit $T_K^{(1)} \ll D$.

\begin{figure}
\centerline{
\includegraphics[width=2.8in]{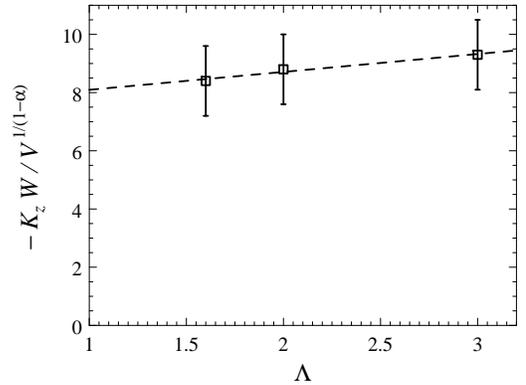}
}
\caption{
$\Lambda$ dependence of the slope of the phase boundary
near the Toulouse point. ($\Lambda$ is the NRG
parameter defining the logarithmic discretization of the
conduction band.)
The dashed line is a linear fit. Each data point involves
an extrapolation of the numerical results at finite negative $W$
to $W\to 0$.
A rather strong $\Lambda$ dependence can be observed, however,
the extrapolated value appears consistent with the analytical
result (\protect\ref{wc2}).
}
\label{fig:slope}
\end{figure}

As mentioned above, some NRG results show a relatively strong dependence
on the NRG discretization parameter $\Lambda$.
Fig.~\ref{fig:nrgphd1} shows the phase diagram for $\Lambda=2$;
results for other $\Lambda$ values are similar, but the $T_K^{(1)}/K_z^{cr}$
values can differ by 50\% or more.
Therefore, we have performed an extrapolation to $\Lambda\to 1$
for a few important quantities.
A sample extrapolation is shown in Fig.~\ref{fig:slope} for the
slope of the phase boundary near the Toulouse point,
which was determined analytically in Sec.~\ref{sec:StrongCoupling} --
the extrapolated value of $K_z W / V^{\frac{1}{1-\alpha}}$ is consistent with
the exact result in Eq. (\ref{wc2}).
We have also looked at the maximum value of $T_K^{(1)}/K_z$ of the
phase boundary occuring near $J_z = 0$, this value
extrapolates to $(T_K^{(1)}/K_z)_{\rm max} = 0.11 \pm 0.03$.

\section{Flow Equations}
\label{sec:Flow}

In this section, we consider a different approach~\cite{kv} to our
original model of Ising-coupled Kondo impurities, which is based on
the method of flow equations.\cite{Wegner94}  The general idea is an
approximate diagonalization of each of the two Kondo impurities -- the
result is a resonant-level-type effective model which captures the
Kondo physics in terms of a non-trivial renormalized hybridization.
Taking the two Ising-coupled impurities together, we will again arrive
at an effective Anderson model.  Away from the Toulouse line one finds
an additional weak density-density interaction.
However, in contrast
to Sec.~\ref{sec:largeW} where we were forced to treat a similar
interaction non-perturbatively to take into account x-ray edge
singularities (App.~\ref{sect:SW}), this is not necessary here because
this non-trivial physics is already contained in the fully renormalized
couplings which naturally appear within the flow-equation approach.
This approach allows for a
systematic expansion around the Toulouse line, and the effective
Hamiltonian derived in this framework in fact describes the entire
phase diagram of Fig.~\ref{fig:phd1} consistently.

\subsection{Flow equation transformation}

The flow equation method was first applied to the Kondo model in
Ref.~\onlinecite{HofstetterKehrein_01}, where it was shown that it
leads to an expansion around the Toulouse point.  Its basic idea is to
perform a sequence of infinitesimal unitary transformations on a given
many-particle Hamiltonian and thereby diagonalizing
it.\cite{Wegner94}
The expansion parameter turns out to be
$\lambda_0-1$ with $\lambda_0=\sqrt{2}(1-J_z \rho)=1-W \rho$ (using
the bosonization cutoff scheme); the Toulouse point corresponds to
$\lambda_0=1$.

Following Ref.~\onlinecite{HofstetterKehrein_01}, we construct unitary
transformations $U_{1,2}$ (see App.~\ref{sect:App_feq}) such that
the single-impurity Hamiltonians $\tilde H_{1,2}^K$ from
Eq.~(\ref{Htilde}) become diagonal \beq H_{1,2}^{(K-{\rm
    diag})}=U_{1,2}^\pdag\,\tilde H_{1,2}^K U_{1,2}^\dag \eeq up to
higher-order terms in our expansion around the Toulouse line.  For
studying the Ising-coupled Kondo impurities we therefore apply the
combined unitary transformation $U=U_1\,U_2$ on (\ref{Model}), $\tilde
H^{K}=U\,H^K\,U^\dag$, leading to \beq \tilde H^K=H_{1}^{(K-{\rm
    diag})}+H_{2}^{(K-{\rm diag})} +K_z\,\tilde S^z_1 \tilde S^z_2
\label{tildeHK}
\eeq
with $\tilde S^z_{1,2}=U_{1,2}^\pdag\, S^z_{1,2}\, U_{1,2}^\dag$.
At the Toulouse point (\ref{tildeHK}) is of course exactly equivalent
to the Anderson impurity model (\ref{RL}) with $W=0$ and the same
mapping as used in Secs.~\ref{sec:Bosonization} and~\ref{sec:GAIM}:
the unitary transformation $U_{1,2}$ just eliminates the hybridization
coupling in the Anderson impurity model.

It was shown in Ref.~\onlinecite{Slezak02} that the flow equation approach
yields a resonant-level model (\ref{RL}) as an effective model for the Kondo
impurity model also away from the Toulouse point
\beq
H^{(RL-{\rm eff})}=H_0[\Psi_{j}]
+ \sum_k \tilde V_k \left(d^\dag_j  \Psi^\pdag_{j}(k) + {\rm h.c.} \right) \ ,
\label{RL_eff}
\eeq
where the $\Psi^\dag_{j}(k), \Psi^\pdag_{j}(k)$ are the creation and
annihilation operators for solitonic spin excitations in momentum space.
However, this resonant-level model now has a nontrivial renormalized hybridization
function,
$\Delta(\epsilon)\stackrel{\rm def}{=}\sum_k \tilde V_k^2
\delta(\epsilon-\epsilon_k)$,
with (i) $\Delta(0)= T_K^{(1)}/w\pi^2$ and nearly constant in an energy
interval of order $T_K^{(1)}$ around the Fermi energy (here $\epsilon_F=0$),
and (ii) a non-trivial power-law behavior for larger energies.
Furthermore, it was shown in Ref.~\onlinecite{Slezak02} that to {\em leading} order
in an expansion around the Toulouse line one can identify
$S_j^z=d^\dag_j d^\pdag_j-1/2$. The effective model for our system
of Ising-coupled Kondo impurities is therefore an Anderson impurity
model with a hybridization function of order the single-impurity Kondo
scale:
\beq
H^{(A-{\rm eff})} = H_0[\Psi_{\sigma}] + \sum_{k,\sigma} \tilde V_k
\left(d^\dag_\sigma \Psi^\pdag_{\sigma}(k) + {\rm h.c.} \right)
+ K_z \bar{n}_{d\uparrow} \bar{n}_{d\downarrow} \ .
\label{H_Aeff}
\eeq
The main feature of the flow equation method is therefore to eliminate
the large coupling~$W$ in (\ref{Anderson}) by renormalizing the
hybridization of the Anderson model.

However, since the flow equation
transformation is an expansion in the distance $(\lambda_0-1)$ to
the Toulouse line, we need to be careful in the transformation
of $S^z$: The transformed $\tilde S^z$ is multiplied by
a possibly large parameter~$K_z$, so that an error of order
$(\lambda_0-1)$ in the expansion becomes multiplied by~$K_z$,
leading to additional interaction terms in (\ref{H_Aeff}) that
can be larger than the hybridization energy scale~$T_K^{(1)}$.\cite{kv}
It are precisely these additional interactions that drive the
Kosterlitz--Thouless transition between the Fermi-liquid phase
and the frozen mini-domain phase for
$|K_z (\lambda_0-1)|\gtrsim T_K^{(1)}$.

\subsection{Corrections to the transformation of $\tilde S^z$}
\label{sec:corrfeq}

In App.~\ref{sect:App_feq} the flow equation solution for the
single impurity Kondo model $H^K_{1/2}$ in a magnetic field~$h$ is
discussed with a careful analysis of terms of order~$(\lambda_0-1)$.
Here $h$ is the effective exchange field due to the
second spin, to be described below.
 The transformed operator $\tilde S^z_j$ takes the
following form \bea \tilde S^z_j &=&\frac{1}{2} \int {\rm d}x\,{\rm d}x'\:
d(x)\,d^*(x')\: [\Psi^\dag_j(x),\Psi^\pdag_j(x')]
  \nn \\
&&+\frac{1}{2}(\lambda_0-1) f(h) \partial_x \bar\phi_j(0)\label{corr_trfsigmaz}\eea
plus irrelevant terms (containing e.g.\ higher derivatives of the
bosonic field) and plus higher order terms of order $(\lambda_0-1)^2$.
The first term on the right-hand side  of Eq.~(\ref{corr_trfsigmaz})
can be interpreted as the result of integrating out the hybridization
term in (\ref{RL_eff}), while the second term is a correction term not
contained in the original solution in
Ref.~\onlinecite{HofstetterKehrein_01}.  $\bar\phi_j(0)$ denotes the
bosonic spin-density field $\phi_j(x)$ without the Fourier components
for energies larger than ${\cal O}(T_K^{(1)})$ (with respect to
low-energy properties one does not need to distinguish these fields).
Pro\-per\-ties of the dimensionless function $f(h)$ are derived in
App.~\ref{sect:App_feq}, in particular $f(h)=v_F/|h|+{\cal O}(h^{-2})$
for $|h|\gg T_K^{(1)}$.

In the coupled system (\ref{H_Aeff}) we can approximate
the effect of one spin on the other as a static magnetic field
of strength $h=\pm K_z/2$
close to the transition.
This approximation becomes asymptotically exact
as one approaches the transition since the spin dynamics
becomes slower and slower.

\begin{widetext}
We arrive at the following
Hamiltonian describing the coupled Kondo impurities
in the vicinity of the transition line.
\bea
H^{(A-{\rm eff})} \approx H_0[\Psi_{\sigma}] + \sum_{k,\sigma} \tilde V_k
\left(d^\dag_\sigma \Psi^\pdag_{\sigma}(k) + {\rm h.c.} \right) + K_z
\bar{n}_{d\uparrow} \bar{n}_{d\downarrow} +(\lambda_0-1) \frac{K_z}{2} f(K_z/2)
\left(\partial_x\bar\phi_\uparrow(0) \bar{n}_{d\downarrow}
+\bar{n}_{d\uparrow} \partial_x\bar\phi_\downarrow(0) \right)
\label{H_Aeffcorr}
\eea
up to corrections of order $(\lambda_0-1)^2$.

\subsection{The Kosterlitz--Thouless transition}\label{flowKT}

Let us now focus on the case $K_z\gg T_K^{(1)}$ that is
relevant for studying the phase transition in the vicinity
of the Toulouse line. Using $f(h)\approx v_F/|h|$ we rewrite the Hamiltonian (\ref{H_Aeffcorr}) as
\bea
H^{(A-{\rm eff})}&=&H_0[\Psi_{\sigma}] + \sum_{k,\sigma} \tilde V_k
\left(d^\dag_\sigma \Psi^\pdag_{\sigma}(k) + {\rm h.c.} \right) + K_z
\bar{n}_{d\uparrow} \bar{n}_{d\downarrow}\nn \\
&& +(\lambda_0-1) v_F
\big( (\partial_x\bar\phi_{\uparrow}(0)
+\partial_x\bar\phi_{\downarrow}(0))
\frac{1}{2}(\bar{n}_{d\uparrow}+\bar{n}_{d\downarrow}) -
(\partial_x\bar\phi_{\uparrow}(0)
-\partial_x\bar\phi_{\downarrow}(0))
\frac{1}{2}(\bar{n}_{d\uparrow}-\bar{n}_{d\downarrow}) \big) \ .
\label{H_Aeffcorr2}
\eea
\end{widetext}
Similar to the analysis in Sec.~\ref{sec:largeK} the term proportional
to $\bar{n}_{d\uparrow}+\bar{n}_{d\downarrow}$ is frozen out and can
be ignored, while the term proportional to
$\bar{n}_{d\uparrow}-\bar{n}_{d\downarrow}$ leads to a spin--spin
interaction. Since $K_z$ is the largest energy scale in
(\ref{H_Aeffcorr2}) with its renormalized parameters, we can map the
Hamiltonian onto an anisotropic Kondo model using a Schrieffer--Wolff
transformation (like in Sec.~\ref{sec:largeK}) and again arrive at
(\ref{effKondo})
\beq H^{A}_{\rm eff} = H_0[\Psi_{\sigma}] +
\sum_{n\alpha\beta} J^{\text{eff}}_{n} S^n \Psi^\dag_{\alpha}(0)
\sigma^n_{\alpha\beta} \Psi^\pdag_{\beta}(0)\ , \label{effKondo2}\eeq
where now the
Kondo couplings for scattering processes in the vicinity of the Fermi
surface contain the renormalized parameters $\rho_F \tilde
V^2=T_K^{(1)}/w\pi^2$ \beq \rho_F J^{\rm eff}_\perp = \frac{4\rho_F
  \tilde V^2}{K_z}\ ,\qquad \rho_F J^{\rm eff}_z = \frac{4\rho_F
  \tilde V^2}{K_z}-\rho_F J^{\rm (nl)} \eeq with $\rho_F J^{\rm
  (nl)}=\lambda_0-1$.  We stress that here it was {\em not} necessary
to use the generalized Schrieffer--Wolff transformation derived in
App.~\ref{sect:SW} as the parameters in (\ref{H_Aeffcorr2}) are
already renormalized due to the flow equation procedure and the
interactions $\propto \lambda_0-1$ are only effective at low energies.
The additional spin--spin interaction $J^{\rm (nl)}$ is ferromagnetic
for couplings to the right-hand side of the Toulouse line
$\lambda_0>1$. This leads to a critical coupling for the
Kosterlitz--Thouless transition to the frozen mini-domain phase \beq
K_z^{cr}=\frac{8\rho_F \tilde V^2}{\lambda_0-1}=\frac{8\tilde V^2}{-W}
\label{feq:Kzcr}
\eeq
or using Eq.~(\ref{tkdef})
\beq
\rho_F K_z^{cr}=\frac{8}{w\pi^2} \frac{T_K^{(1)}}{-W}
=1.964 \frac{T_K^{(1)}}{-W}
\eeq
in exact agreement with the NRG results (Fig.~\ref{fig:slope}) and the strong-coupling analysis Eq.~(\ref{wc3})
in Sec.~\ref{sec:largeW}.

\begin{figure}
\centerline{
\includegraphics[width=3.1in]{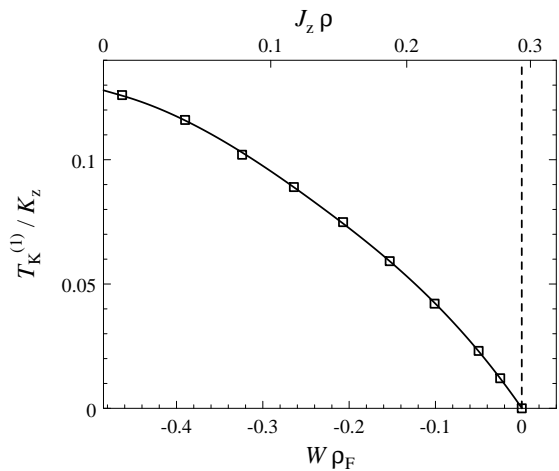}
}
\caption{
Phase diagram of the generalized single-impurity
Anderson model (\protect\ref{Anderson}), deduced using
the flow equation method.
Notation is as in Fig.~\protect\ref{fig:nrgphd1}.
}
\label{fig:feq}
\end{figure}

Since the flow equation approach leads to a renormalized
effective Hamiltonian, one can also use it to derive the
entire phase diagram like in Fig. \ref{fig:nrgphd1}. If one
neglects the same higher order terms $(\lambda_0-1)^2$
as before, one finds the following result for the critical
coupling
\beq
\rho_F K_z^{cr}=\frac{8}{w\pi^2} \frac{T_K^{(1)}}{-W} \:
(1-\Lambda'(0)) \ ,
\label{feq_entKT}
\eeq
which simply results from a Schrieffer-Wolff transformation of
an Anderson model with an on-site repulsion $K_z$ and
a hybridization $\Delta(\epsilon)$ which enters as $T_K^{(1)}$.
Here
\beq
\Lambda(\omega)=\frac{1}{\pi} P\int d\epsilon \,
\frac{\Delta(\epsilon)}{\omega-\epsilon}
\eeq
follows from the effective hybridization function of the
flow equation approach \cite{Slezak02}. The factor
$(1-\Lambda'(0))$ enters in (\ref{feq_entKT}) because
it generalizes the relation between the Kondo
temperature defined in  (\ref{tkdef}) and the renormalized
hybridization function within the effective resonant
level model \cite{Slezak02}
\beq
\pi \Delta(0) = \frac{T_K}{w}\,(1-\Lambda'(0)) \ .
\eeq
The results are depicted in Fig.~\ref{fig:feq}.
The maximum value of the phase boundary occuring near
$J_z=0$ is given by $(T_K^{(1)}/K_z)_{\rm max}=0.126$,
which agrees with the extrapolated NRG value $0.11 \pm 0.03$
from Sect.~\ref{sec:NRG:boundary}.


\section{Symmetries and Perturbations}
\label{sec:Pert}

To what extent do the results presented in the previous sections
depend on the details of the models under consideration? To answer
this question we will investigate whether and how  (small) perturbations
of (\ref{Model}) will qualitatively change the physics.
Fermi-liquid phases with vanishing residual entropy $S_0$ are stable
against small perturbations. This is not necessarily the case for our
``frozen mini-domain'' characterized by $S_0=\ln 2$. The existence of
this $S_0=\ln 2$ phase is a fundamental feature of our model (\ref{Model}).
The necessary conditions for its stability will be discussed in what follows.

Firstly, let us consider the effect of a magnetic field in
$z$-direction acting on the impurity spins. A staggered magnetic
field, $h_s (S^z_1-S^z_2)$, will directly destroy the degeneracy of
the two configurations $|\!\uparrow\downarrow\rangle$,
$|\!\downarrow\uparrow\rangle$. However, in the limit $K_z \to \infty$
a homogeneous magnetic field $h (S^z_1+S^z_2)$ will not destroy the
$S_0=\ln 2$ phase.  It is interesting how these terms modify the
generalized Anderson model (\ref{Anderson}).  The magnetic field $h$
results in a term $h \sum_\sigma d^\dag_\sigma d^\pdag_\sigma $ which
breaks particle--hole symmetry in the generalized Anderson model. It
therefore modifies only the position of the phase boundary.  The
staggered magnetic field $h_s$, however, will lead to a term $h_s
\sum_\sigma \sigma d^\dag_\sigma d^\pdag_\sigma$ which corresponds to
a (pseudo-) magnetic field acting on the pseudospin of the Anderson
model. Only the staggered magnetic field is a relevant perturbation
which destroys the $\ln 2$ phase.

Apart from these magnetic fields in $z$-direction there are other relevant
terms which lift the twofold degeneracy,
which are of the forms:
\bea
S^+_j && j=1,2, \label{transM}\\
 S^+_1 S^-_{2}\,, && \label{transCoup}\\
 S^+_1 S^-_{2} \Psi^\dag_{i \sigma} \Psi^\pdag_{j \sigma} && i,j = 1,2, \label{Op2}\\
S^+_1 S^-_{2} \Psi^\dag_{i \alpha} {\boldsymbol \sigma}_{\alpha \beta}
\Psi^\pdag_{j \beta} && i,j = 1,2 \label{Op3}
\eea
and their hermitian conjugates. It turns out that all these operators
are forbidden if we impose the following two symmetry conditions:
The model should be invariant under the two {\em separate} spin rotations
of each impurity
and its electronic bath about an angle of $\pi$, i.e., under the
transformation
\begin{equation}
U_j = {\rm e}^{i \pi  I^z_j}
\end{equation}
with $j=1,2$. $I^z_j$ is the $z$-component of spin of
system $j$, $I^z_j = S^z_j + \sum_{k} c^\dag_{k \alpha j}
\frac{1}{2}\sigma^z_{\alpha\beta} c^\pdag_{k \beta j}$. In the
presence of these $\pi$ rotation symmetries, $U_j$, the terms
(\ref{transM}), (\ref{transCoup}), (\ref{Op2}) and (\ref{Op3}) are
absent and the frozen mini-domain phase survives. The quantum phase
transition from the frozen mini-domain with residual entropy $\ln 2$
to the phase of Kondo screened impurities therefore just relies (in
the absence of a staggered magnetic field) on the symmetries $U_1$ and
$U_2$.

The model (\ref{Model}) considered in this paper possesses by
construction symmetries beyond $U_j$. They are not necessary for the
stability of the $S_0=\ln 2$ phase.  For example, the two baths are assumed to have
the same Kondo coupling $J_{n}$. This parity
symmetry can be relaxed without destroying the frozen mini-domain
phase. Furthermore, the $z$-component of spin of each system, $I^z_j$,
is conserved in our model since we chose $J_x = J_y = J_\perp$. This
symmetry can also be perturbed without lifting the twofold degeneracy.
Moreover, the frozen mini-domain phase is stable against breaking of
particle--hole symmetry which we implicitly assumed in the
bosonization treatment by linearizing the dispersion relation of the
conduction electrons.
In all these situations, we therefore expect that all of the {\em qualitative} results, i.e.,
the structure of the phase diagram and the nature of the quantum phase transition,
are not affected.

However, any perturbation which breaks either  $U_1$ or  $U_2$ (or both) will
generically generate one of the relevant couplings (\ref{transM}--\ref{Op3})
which all destroy the $\ln 2$ phase.
In the following we briefly discuss two such cases which are likely to occur
in experimental realizations (a third case, corresponding to (\ref{Op2}) is
studied in Sec.~\ref{capacity}).

First, consider a situation where a small spin-flip coupling (\ref{transCoup})
is added on top of the large Ising interaction of the spins,
\bea \label{TransverseCoupling}
\delta H^{\perp}_{12} = K_{\perp} \left(S_1^x S_2^x +S_1^y S_2^y\right)\,.
\eea
In realizations of our model based on spins and strongly anisotropic
spin--orbit interactions, such a term will always be present.
A small $K_{\perp}$ will immediately lead to a tunneling
between the two states of our mini-domain: their degeneracy is lifted,
the two spins form a singlet and the $\ln 2$ residual entropy is
quenched completely.
Two-impurity Kondo models with $K_\perp=K_z$ have been widely
studied.\cite{2imp,2impnrg,2impsakai,2impfye,2impcft,2impPhase,2impbos,2impoli}
As argued in
Refs.~\onlinecite{2impPhase,2impcft} the resulting phase diagram
depends on the presence or absence of particle--hole symmetry
(which does, however, {\em not} modify the phase diagram for $K_\perp=0$
as pointed out above).
In the absence of particle--hole symmetry, the phase transition is replaced by
a smooth crossover. However, in the presence of particle--hole
symmetry, the scattering phase shifts of the electrons can only take
the values $0$ or $\pi/2$. As the Kondo-screened phase and the
inter-impurity singlet phase have different phase shifts, there has
to be a phase transition in between. This transition is {\em not} of
Kosterlitz--Thouless type, but characterized\cite{2impcft,2impbos} by a
residual entropy $\ln \sqrt{2}$. Nevertheless, this transition will
merge with ours in the limit $K_\perp \to 0$, as an infinitesimal
$K_\perp$ does not affect the Kondo-screened phase but leads
immediately to the formation of an inter-impurity singlet in the
frozen mini-domain phase.

A second interesting case is a situation where the two Fermi seas are coupled,
e.g., by a tunneling between the two leads
\bea \label{TunnelingTerm}
\delta H^{\rm tunneling}_{12} =
\sum_{k, k', \alpha}
\left( t_{k k'} :c^\dag_{k \alpha 1} c^\pdag_{k' \alpha 2}: +\: {\rm h.c.} \right)\,.
\eea
While this term is not relevant by power counting, it will induce an
RKKY interaction between the spins and therefore generate the relevant
coupling (\ref{transCoup}) or (\ref{TransverseCoupling}).
As such a term also breaks particle--hole symmetry, the quantum
phase transition will be replaced by a smooth
crossover.


\section{Transport}
\label{sec:Appl}

In this section we illustrate how the phase diagram and, more
importantly, the corresponding quantum phase transition can be revealed
in transport experiments.
We shall discuss two experimental setups:
(A) Capacitively coupled quantum dots, where the charge degrees of freedom
play the role of (pseudo)spins,\cite{Matveev91,Matveev95}
are a promising realization of our model.
By adding a small inter-dot tunneling term, we obtain
a characteristic zero-bias anomaly.
(B) If the Ising coupling is realized between real spin
degrees of freedom, then we shall show that a transport experiment
can reveal a universal fractional critical conductance at the phase transition
which is related to the universal jump of the superfluid density
at the Kosterlitz-Thouless transition of superfluid thin films.
Using quantum dots this situation is difficult to achieve,
as a transverse spin coupling will always be present in the experiment.
Nevertheless, the following proposals highlight the non-trivial
effects of a Kosterlitz-Thouless transition with a bath of solitonic particles
onto the original electrons.

\subsection{Zero-bias anomaly of capacitively coupled quantum dots}\label{capacity}

\begin{figure}
\centerline{
\psfrag{V}{$V$}
\psfrag{VG1}{$V_{G1}$}
\psfrag{VG2}{$V_{G2}$}
\psfrag{t}{$t$}
\psfrag{Jp}{$J_\perp$}
\includegraphics[width=1.2in]{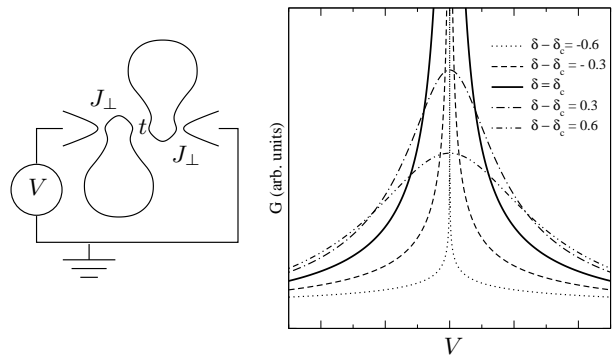}
\hspace{5pt}
\includegraphics[width=1.8in,clip=]{matveevG.eps}
}
\caption{Left panel:
  Experimental setup to measure the tunneling conductance between two
  capacitively coupled quantum dots. Right panel: Schematic plot of the zero
  bias anomaly of the conductance at $T=0$.  In the ``frozen
  mini-domain'' phase, $\delta<\delta_c$, the conductance diverges
  algebraically according to Eqn.~(\ref{gFrozenV}). At the quantum
  phase transition, $\delta=\delta_c$, the exponent takes the
  universal value $-2(\sqrt{2}-1)$ according to
  Eq.~(\ref{gMatveevCr}). In the Kondo screened phase,
  $\delta>\delta_c$, the conductance is finite for $V\to 0$.
\label{fig:MatveevSetup}}
\end{figure}

In realization of our model (\ref{Model}) using {\em charge}
states of capacitively coupled quantum dots\cite{Matveev91}
the conductance discussed above cannot be easily measured.
We therefore propose
another experiment sketched in Fig.~\ref{fig:MatveevSetup}.  We consider two
large quantum dots, each coupled to a (single-channel) lead.
The Coulomb interaction
and an appropriately choosen gate voltage ensure that two charge states
on each dot are degenerate, and that all other charge states have
higher energies.
These two charge states in each dot take over the
role of the two spins as explained in more detail by
Matveev.\cite{Matveev91}
The spin up and down states of the conduction
electrons in our model correspond to electrons sitting either in the
leads or on the dot, where we assume that the level spacing on this dot
is small compared to temperature. Using this mapping, a capacitive
coupling of the two dots directly corresponds to an Ising coupling
(\ref{IsingCoupling}). The physical spin in such a system would
translate to an extra channel index in our model.
For simplicity we will, however, consider a situation where either strong
spin--orbit scattering mixes those channels or where the spin is quenched by a
strong magnetic field -- in both cases we effectively deal with spinless
fermions and thus with a single-channel model.

We now consider a situation where the two dots are coupled by weak
tunneling $t$ in addition to the large inter-dot capacity.
This tunneling takes the form
\begin{eqnarray} \label{InterdotTunneling}
H_{\rm tun}
&=& t  S^+_1 S^-_2
\Psi^\dag_{\downarrow 1}(0) \Psi^\pdag_{\downarrow 2}(0) + {\rm h.c.}
\end{eqnarray}
We assume that the tunneling is sufficiently weak that it can be
treated perturbatively in the experimentally relevant temperature
range. This is precisely the situation which was also considered by
Andrei {\em et al.}\cite{Andrei}
Note that the approximation to consider only tunneling
into $\Psi_{\sigma i}(x=0)$ in Eq.~(\ref{InterdotTunneling}) is only
valid if the tunnel contact between the dots is sufficiently close to
the lead contact (see Fig.~\ref{fig:MatveevSetup}).

We calculate the conductance in perturbation theory in the
interdot tunneling $t$ starting from the Kubo formula. The current
through the link between the dots is then given by $j= t S^+_1 S^-_2 (i
\Psi^\dag_{\downarrow 1}(0)\Psi^\pdag_{\downarrow 2}(0)) +{\rm h.c.}$
and the $T$ dependence of the current--current correlator can be
obtained from simple power counting.

We first consider the ``frozen mini-domain'' phase.  Following the
arguments given in Sec.~\ref{sec:StrongCoupling}, the dimension of
the tunneling term (or equivalently of the current operator) with
respect to this fixed point is given by
$\text{dim}[H_\text{tun}]=\text{dim}[j] =
1-(\frac{2\delta}{\pi})^2-(1-\frac{2\delta}{\pi})^2$. Therefore the
current--current correlator decays in time as $t^{-2
  (\text{dim}[j]-1)}$, and we obtain for the conductance
\begin{equation}\label{gFrozenT}
G(T)\sim t^2 T^{-2 \text{dim}[j]}=t^2
T^{-4 \frac{2\delta}{\pi} \left(1-\frac{2\delta}{\pi}\right)}\,.
\end{equation}
This divergence of the conductance arises because the tunneling is a {\em
  relevant} perturbation which will finally destroy the ``frozen
mini-domain'' phase and quench its residual entropy $\ln 2$ below some
small energy scale.  Eq.~(\ref{gFrozenT}) is therefore only valid for
sufficiently small $t$ when this scale is smaller than $T$.
Furthermore, a finite domain-flip rate induced by (\ref{DomainFlip})
is required to obtain a finite current. Above we implicitly assumed
that $t$ is so small that the size of the current is solely determined
by the smallest bottleneck for charge transport given by $t$.

At finite voltage $V\gg T$, $T$ in (\ref{gFrozenT}) can be replaced by
$V$ and we expect a  zero-bias anomaly characterized by a pronounced
peak in the conductance:
\begin{equation}\label{gFrozenV}
G(V)\sim  |V|^{-4 \frac{2\delta}{\pi} \left(1-\frac{2\delta}{\pi}\right)}\,.
\end{equation}
Upon approaching the quantum phase transition, the divergence
increases and at the KT transition takes
the universal form
\begin{eqnarray}\label{gMatveevCr}
G_{cr}(T)&\sim&
T^{-2(\sqrt{2}-1)}\approx T^{-0.83} \,, \\
G_{cr}(V)&\sim&
|V|^{-2(\sqrt{2}-1)}\approx |V|^{-0.83}
\end{eqnarray}
up to logarithmic corrections.

In the Kondo-screened phase, we can calculate the qualitative behavior
of the current--current correlator at the point in the phase diagram
where $K_z=0$ and the dots decouple. The current--current correlator
then can be decomposed into two correlators of the form $\left\langle
  S^+_i(t) \Psi^\dagger_{\downarrow i}(t) S^-_i(0)
  \Psi^\pdag_{\downarrow i}(0)\right\rangle$ which decay
asymptotically as $1/t$. This can be seen if one identifies this
correlator with the conduction electron T matrix (see
Ref.~[\onlinecite{roschSpectral}] and references therein) which is
characterized by a {\em constant} spectral density for low energies.
The conductance therefore approaches a constant for temperatures and
voltages well below the characteristic crossover temperature $T^*$ to
the Kondo-screened phase:
\begin{eqnarray}\label{gsing}
G(V)\approx G(T)\approx {\rm const.}
\end{eqnarray}
In Fig.~\ref{fig:MatveevSetup} we show schematically the nonlinear
conductance as a function of $V$ in the vicinity of the quantum phase
transition.

In contrast to Eq.~(\ref{gFrozenT}) and Eq.~(\ref{gsing}),
N.~Andrei {\it et al.~}\cite{Andrei} obtained an exponentially small
conductance in the ``frozen mini-domain'' phase and $G\sim T^4$ in the singlet
phase, with which we disagree.

\subsection{Universal conductance of Ising-coupled quantum dots}

What is the most characteristic signature of the Kosterlitz--Thouless
quantum phase transition which we found in the previous sections?
The most famous example of a Kosterlitz--Thouless transition is probably the vortex
binding--unbinding transition in superfluid $^4$He films.
From the Kosterlitz--Thouless
theory follows the prediction of an universal jump in the superfluid
density upon passing through the transition.\cite{jump}

Interestingly, the analogue of the superfluid density in our model is
the scattering phase shift $\delta$ of the conduction electrons, and
the arguments for an universal jump in the superfluid density carry
over to an universal jump in $\delta$.
This can be seen by considering
the RG flow diagram Fig.~\ref{fig:kondoflow}: In the ``frozen
mini-domain'' phase the system flows towards a {\em line} of fixed
points which is  naturally characterized by the dimension of the
leading irrelevant operator, i.e. the domain flip (\ref{DomainFlip}),
or, equivalently, according to (\ref{dimHflip}) by the phase shift
$\delta$ of the conduction electrons. Upon approaching the quantum
phase transition, the irrelevant domain flips become marginal and the
phase shift increases and reaches $\delta_T=\frac{\pi}{2}
\left(1 - \frac{1}{\sqrt{2}}\right)$ at the phase boundary [see
Eq.~(\ref{deltaTdef})]. On the other side of the phase diagram, the
system flows to the strong-coupling fixed {\em point}, where the Kondo
spins are screened and the electrons acquire a phase shift of $\pi/2$.
Therefore the phase shift jumps across the transition from $\delta_T$
to $\pi/2$!
This picture is expected to hold everywhere close to the
phase boundary as long as no other phase transition intervenes -- that
the latter does not happen is shown by our NRG calculations.
\begin{figure}
\centerline{
\psfrag{G}{$G$}
\psfrag{G1}{$G_0$}
\psfrag{G2}{$G_{cr}$}
\psfrag{K}{$\delta$}
\psfrag{Kcr}{$\delta_T$}
\psfrag{a}{}
\psfrag{b}{}
\psfrag{T^2}{$T^2$}
\psfrag{T^d}{$T^{2 d}$}
\psfrag{log1}{$\frac{1}{\log T}$}
\psfrag{log2}{$\frac{1}{\log^2 T}$}
\psfrag{T0}{$\scriptstyle T=0$}
\psfrag{T!0}{$\scriptstyle T > 0$}
\includegraphics[width=1.4in]{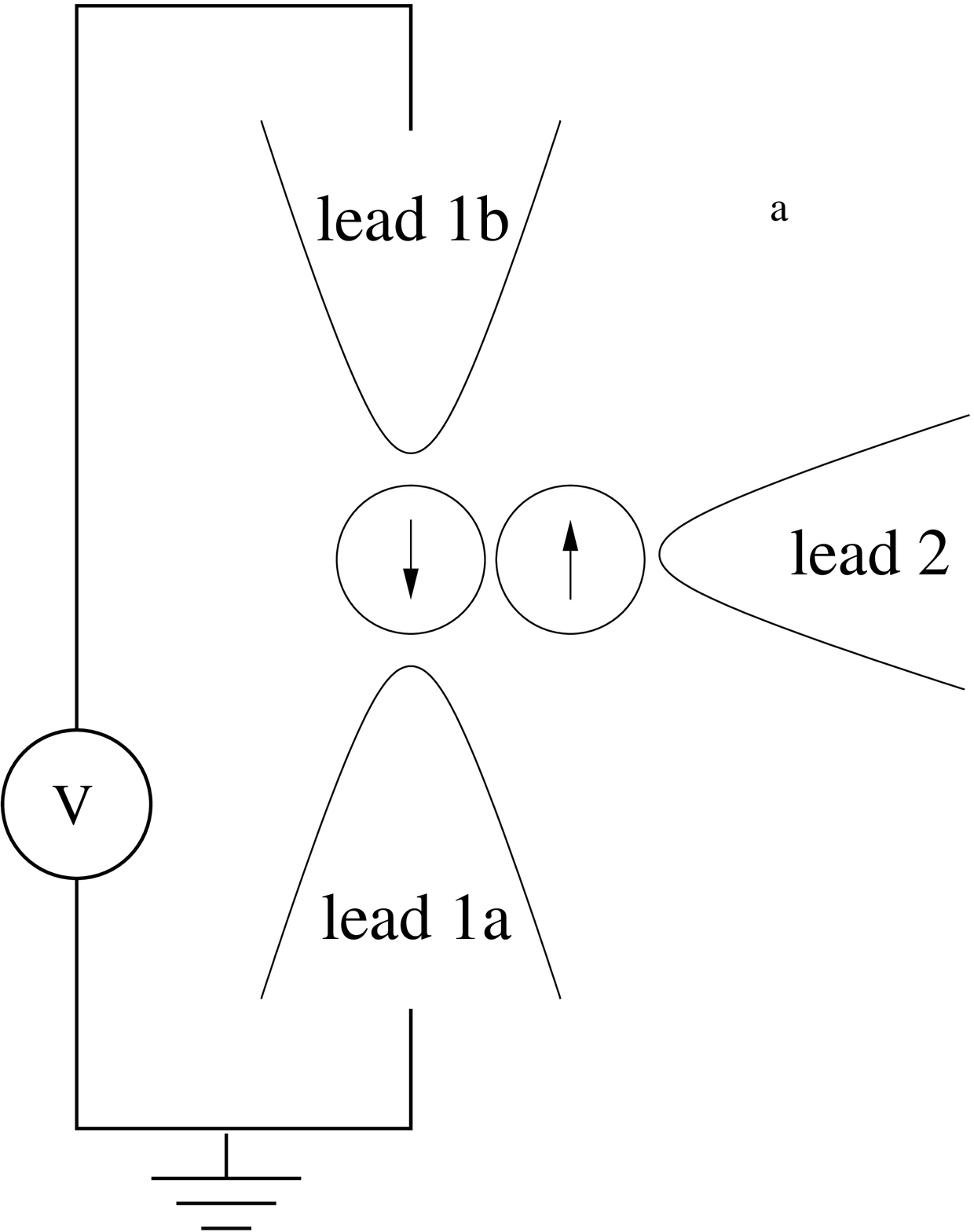}
\hspace{5pt}
\includegraphics[width=1.6in,height=1.8 in]{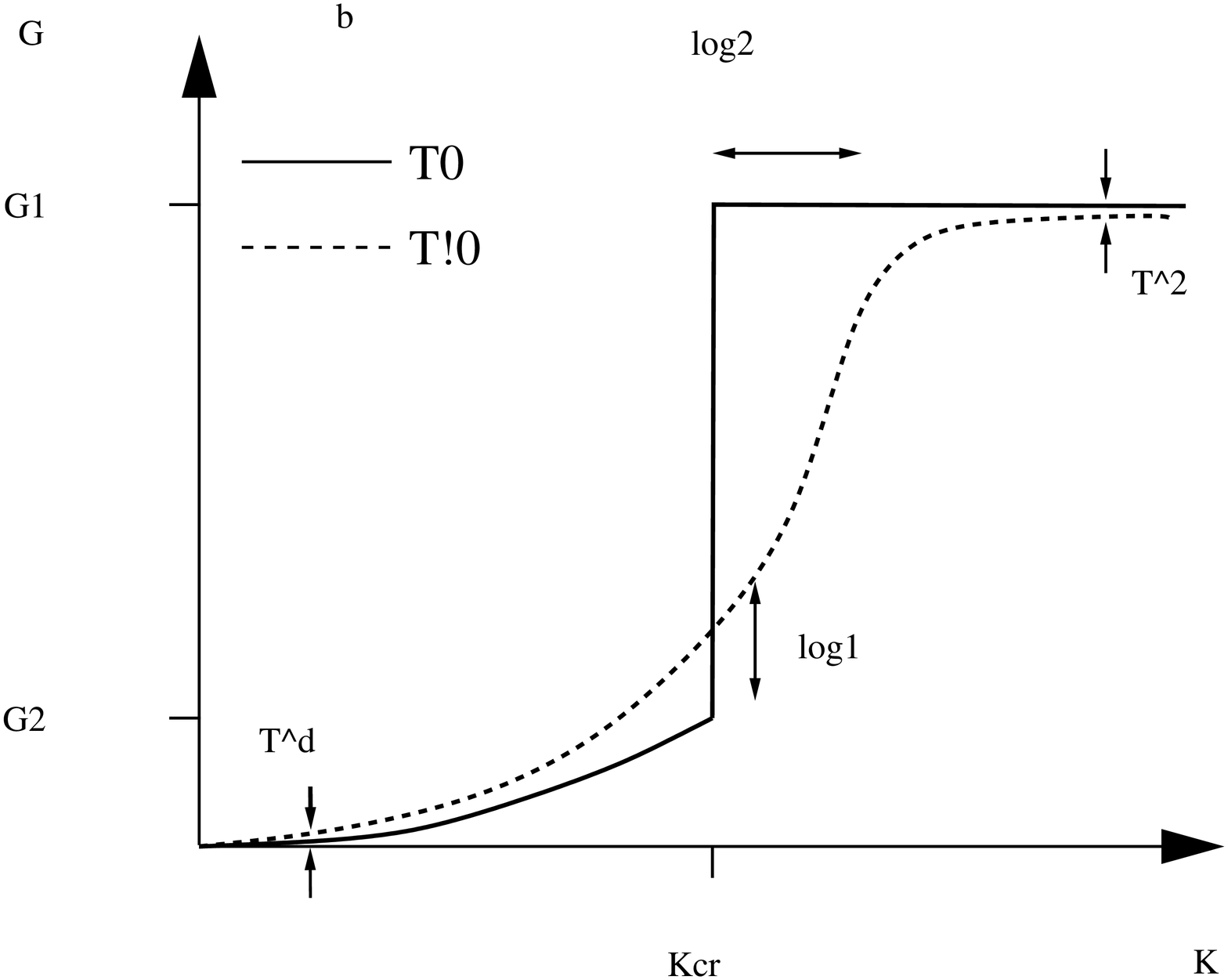}
}
\caption{Left panel: Experimental setup to measure the conductance
  through a single quantum dot, which is Ising-coupled to a second dot.
  The couplings to the leads and between the dots can be tuned using appropriate
  gate voltages.
  Right panel: At $T=0$, the conductance (solid
  line) takes at the quantum phase transition the universal value
 $G_{cr}=G_0 \cos^2 \frac{\pi}{2 \sqrt{2}}$,  (\ref{univCritCond}).
  Dashed line: schematic plot of the conductance at finite $T$.
  Corrections to the $T=0$ result are logarithmic at the transition.
  The exponent $2 d \equiv -2 {\rm dim} [H^{\rm flip}_{\rm eff}] $ is
  given by the dimension of the domain flip term (\ref{DomainFlip}).
}
\label{fig:expSetup}
\end{figure}

This universal jump of the phase shift has direct experimental
consequences. Consider the experimental setup sketched in
Fig.~\ref{fig:expSetup} where the conductance through the left dot is
measured.  If Kondo screening prevails, the conductance for $T\to0$
will be given by the conductance quantum $G_0=2 e^2/(2 \pi \hbar)$.
In the frozen mini-domain phase on the other side of the phase
diagram, spin flips are completely supressed for $T\to 0$ and
therefore we can assume a {\em static} spin configuration to calculate
$G(T=0)$.  For such a potential scattering problem, the conductance is
given by
\begin{eqnarray}\label{univCond}
G(T=0)=G_0 \sin^2 \delta\,.
\end{eqnarray}
Directly at the quantum phase transition, the  conductance therefore takes the universal value
\begin{multline}\label{univCritCond}
G_{cr}(T=0)=G_0 \sin^2 \delta_T=G_0 \cos^2\left[\frac{\pi}{2 \sqrt{2}}\right]\approx 0.2 \, G_0 \,,
\end{multline}
and it jumps to the Kondo value $G_0$ upon entering the Kondo-screened
phase.
This universal fractional conductance at our quantum phase
transition is one of the remarkable results of this paper.

It is interesting to compare this to the well-known result for the
usual Kondo effect, where the conductance jumps from $0$ to $G_0$ when
the exchange coupling $J$ is tuned from ferromagnetic to
antiferromagnetic. Both in Sec.~\ref{sec:largeW} and Sec.~\ref{flowKT}
we mapped our model close to the quantum phase transition to such a
Kondo model. The fermionic degrees of freedom in these Kondo models
[Eq.~(\ref{effKondo}) or Eq.~(\ref{effKondo2})] are, however, complex
{\em solitonic} excitations in terms of the original fermions. While
the phase shifts of those solitons vanishes at the quantum phase
transition, the phase shift of the {\em physical} electrons takes the
fractional value $\delta_T$ leading to a fractional conductance.
Also in other systems which are described by a Kosterlitz--Thouless quantum
transition in terms of solitons, a universal fractional conductance
of similar origin can be expected at the transition.

In Fig.~\ref{fig:expSetup} the zero-temperature conductance close to the phase
transition is shown. At any finite temperatures, the jump in the
conductance is strongly smeared as sketched schematically in the figure.
The $T$-dependence at lowest temperature is determined by the
dimension of the leading irrelevant operators. In the Kondo-screened
phase leading corrections for $T\to 0$ to the Kondo conductance $G_0$
are of order $(T/T^*)^2$ for $T\ll T^*$ , where $T^*$ is exponentially
small close to the quantum phase transition (see Fig.~\ref{fig:kzfit}). In the ``frozen
mini-domain'' phase, corrections to (\ref{univCond}) vanish as
$T^{-2 {\rm dim} [H^{\rm flip}_{\rm eff}] }$ where
${\rm dim} [H^{\rm flip}_{\rm eff}]$
is defined in Eq.~\ref{dimHflip}. Directly at the
quantum phase transition the exponent vanishes and leading corrections to
Eq.~(\ref{univCritCond}) are of the order $1/\ln T$ and therefore
rather large.


\section{Summary}

We have investigated a model of two Ising-coupled Kondo impurities
using strong-coupling expansion, numerical renormalization group
calculations, and a transformation based on the method of flow
equations.  Those methods yield consistent results and allowed us to
show the existence of a Kosterlitz--Thouless phase transition between a
Fermi-liquid phase and a pseudospin doublet phase which corresponds to
a ``frozen mini-domain''.  This transition can be tuned both by
varying the Ising coupling between the impurities and by varying the
anisotropy of the individual Kondo couplings.  In particular, at the
Toulouse point of the individual Kondo impurities we could map the
model {\em exactly} to an Anderson impurity model with a Fermi sea
consisting of fermionic soliton excitations -- in this situation no
phase transition occurs, and the system is in the Fermi-liquid phase,
where the impurity pseudospin is screened below a collective Kondo
scale $T^\ast$.  For $J_z$ smaller than the Toulouse point value,
large $K_z$ drives the system into the pseudospin doublet phase.

The most promising way to realize our model is the situation
of capacitively coupled
quantum dots where the impurity spins represent charge degrees
of freedom on the dots. We have shown that a small additional
tunneling between the dots gives rise to a zero-bias conductance
anomaly with a {\em universal} fractional power-law occurring at the
transition point.
In addition, we have discussed a setup which is interesting
on theoretical grounds, namely transport through one quantum dot of a pair of
dots with a magnetic Ising coupling, where we have found a {\em universal}
fractional conductance through the device at the phase transition point.

\begin{figure}
\centerline{
\psfrag{log 4}{$S\approx\ln 4$}
\psfrag{T0}{$T_0$}
\psfrag{T*}{$T^*$}
\psfrag{log 2 + c1}{$S\approx \ln 2 \! + \! {\cal O}(\ln T)$}
\psfrag{log 2 + c2}{$S\approx \ln 2 + {\cal O}(T^{\alpha})$}
\psfrag{log 1}{$S\approx 0$}
\psfrag{Kz}{$K_z$}
\psfrag{Kzc}{$K_z^{cr}$}
\psfrag{0}{$0$}
\psfrag{T}{$T$}
\psfrag{1}{$T_K^{(1)}$}
\includegraphics[width=2.3in,clip=]{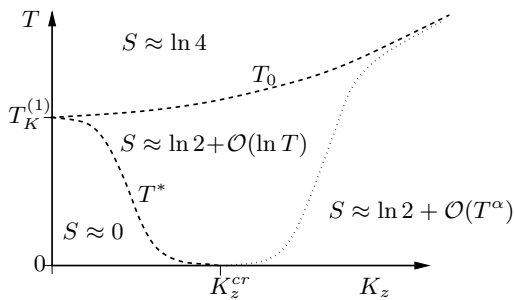}
}
\caption{
  Schematic phase diagram as a function of $K_z$ and $T$ for fixed $T_K^{(1)}$.
  For $T=0$
  there is a quantum phase transition at $K_z=K_z^{cr}$ from a Fermi
  liquid with residual entropy $S_0=0$ to the ``frozen mini-domain''
  phase with $S_0=\ln 2$.  At $T>0$, only smooth crossovers occur
  indicated by the dashed and dotted lines.  At the dashed lines, the
  entropy $S$ changes by $\ln 2$ (see also Fig.~\protect\ref{fig:entr}),
  while at the dotted line one obtains
  a crossover from a logarithmic to a power-law behavior in the
  leading corrections to $S$. Similar crossovers also occur in
  transport quantities. Below $T_0$ a magnetic ``mini-domain'' is
  formed, while a Fermi liquid is recovered below $T^*$ which is
  exponentially small close to $K_z^{cr}$.
\label{phasediagramKT}
}
\end{figure}

With an eye towards comparison with experiments we discuss
the finite-temperature crossover behavior across the phase
diagram (see also Fig.~\ref{fig:entr}.)
If we fix the parameters of the individual Kondo
impurities, then varying $K_z$ corresponds to a vertical
cut through the phase diagram in Fig.~\ref{fig:phd1};
the resulting finite-temperature behavior is sketched
in Fig.~\ref{phasediagramKT}.

For small $K_z$ there is a single crossover at the
single-impurity Kondo temperature $T_K^{(1)}$.
This crossover splits into two when $K_z$ approaches values
of order $T_K^{(1)}$ -- then the described two-stage quenching
of the entropy is observed.
The upper crossover temperature, $T_0$, is associated with the
formation of the magnetic mini-domain, where relative
fluctuations of the two impurity spins are frozen out.
The lower crossover temperature is the collective Kondo scale
$T^\ast$ below which the pseudospin of the mini-domain is
screened.
$T^\ast$ becomes exponentially small near $K_z^{cr}$ and vanishes
for $K_z\geq K_z^{cr}$.
For $K_z\geq K_z^{cr}$ another crossover line appears which, however,
has much weaker signatures, namely the character of the leading
corrections to the entropy and other quantities changes,
as is easily understood from the RG flow in Fig.~\ref{fig:kondoflow}.
For large $K_z$ the entropy change from $\ln 4$ to $\ln 2$ occurs
around $T\sim K_z$, therefore $T_0$ approaches $K_z$ in this limit.

Interestingly, the different impurity degrees of freedom can
be re-interpreted: the flipping of the pseudospin while keeping
the mini-domain intact apparently corresponds to pseudospin ``phase''
fluctuations, whereas breaking up the mini-domain is related to ``amplitude''
fluctuations of the pseudospin.
Thus, in Fig.~\ref{phasediagramKT} we encounter the situation that
amplitude fluctuations are frozen out at a higher temperature $T_0$
whereas phase fluctuations are quenched at the lower $T^\ast$, in other
words, the two impurity spins fluctuate independently for $T>T_0$ whereas
they fluctuate in a correlated fashion between $T^\ast<T<T_0$.
This physics is surprisingly similar to the behavior of lattice
systems in low dimensions, with the difference that of course no
true ordering can occur in the impurity model.

In summary, the present two-impurity model shows remarkably
rich behavior, which awaits realizations in mesoscopic devices.
An interesting extension would be the two-channel case which
is naturally met in capacitively coupled dots with spin-degenerate
conduction electrons.


\acknowledgments

The authors acknowledge fruitful discussions with N.~Andrei,
R.~Bulla, L. I. Glazman, K. Le Hur, and A. Schiller.
This work has been supported by the Deutsche Forschungsgemeinschaft (DFG)
through SFB 484 (SK,TP), the Emmy-Noether program (MG,AR), and the Center for Functional
Nanostructures at the University of Karlsruhe (MV).
SK also acknowledges support by the DFG through a Heisenberg fellowship.


\bigskip

\appendix


\section{Generalized Schrieffer--Wolff transformation}
\label{sect:SW}

In this appendix, we perform explicitely the mapping of the
generalized Anderson model (\ref{Anderson}) to the Kondo Hamiltonian
(\ref{effKondo}) for large $K_z$. Due to the presence of the
interaction $W$ in (\ref{Anderson}) the usual Schrieffer--Wolff
transformation has to be generalized to take into account power-law
renormalizations (of ``x-ray edge'' type) induced by $W$.  We derive
the mapping by investigating directly the properties of a perturbative
expansion in the hybridization $V$ for {\em finite} $W$ within the
Anderson model.  Consider the generalized Anderson model in its
bosonized version. We first eliminate the $W$ term in (\ref{Anderson})
by the Emery--Kivelson transformation (\ref{EK}) with $\gamma^* = W
\rho$ and obtain
\begin{equation}
U^\pdag_{\gamma^*} H^A U^\dagger_{\gamma^*} =
\sum _\sigma H_0[\phi_\sigma] +  K_z \bar{n}_{d\uparrow} \bar{n}_{d\downarrow} +
H_{\rm int}
\end{equation}
where $W$ enters only the hybridization term
\bea
H_{\rm int} = \frac{V}{\sqrt{2 \pi a}} \sum_\sigma
\left(  d^\dag_\sigma {\rm e}^{-i (1-W \rho) \phi_{\sigma}(0)}
F^\pdag_\sigma + {\rm h.c.} \right)\,.
\eea
For large $K_z$ the $d$-level is only singly occupied. $V$ induces
virtual fluctuations to the the doubly occupied and empty state which
are separated from the singly occupied states $|\uparrow\rangle$, $
|\downarrow\rangle$ by an energy $K_z/2$. To derive the effective
Kondo model consider the S-matrix with respect to this low-energy subspace:
\bea
\lefteqn{ T \exp \big[- i \int\limits_{-\infty}^\infty {\rm d}t H_{\rm int}(t) \big]  =}
\nonumber \\
&&\sum\limits_{n=0}^\infty \int\limits_{-\infty}^\infty
\underset
{t_{2 n} >\dots>t_1}{ {\rm d} t_{2 n}\dots {\rm d} t_1}
 i H_{\rm int}(t_{2 n}) \dots  i H_{\rm int}(t_1)
\eea
$H_{\rm int}$ describes processes from the low-energy sector to high
energies or back. Such virtual excitations are rare and exist only for
a short time if $K_z$ is large. Therefore we can group them to pairs
to obtain an effective interaction living in the low-energy Hilbert space,
\begin{widetext}
\begin{eqnarray}
\int\limits_{-\infty}^{t_{2 m +1}}
{\rm d} t_{2 m} \int\limits_{-\infty}^{t_{2 m}} {\rm d} t_{2 m-1}
 i H_{\rm int}( t_{2 m}) i H_{\rm int}( t_{2 m -1})
\,\approx\, - i \int_{-\infty}^{T_{m+1}} H^{\text{eff}}_\text{int}(T_m)\,,
\\
\text{with} \qquad H^{\text{eff}}_\text{int}(T_m)=
- i \int\limits_{0}^{\infty} {\rm d} t\,
  H_{\rm int}(T_{m}+t/2)  H_{\rm int}(T_{m}-t/2)\nonumber\,,
\end{eqnarray}
where we introduced the center-of-time and relative coordinates.
Interactions between adjacent virtual excitations can be neglected to
leading order for large $K_z$.
Introducing the spin notation, $S_z = \frac{1}{2}\sum_\sigma \sigma
d^\dag_\sigma d^\pdag_\sigma$ and $S^+ = d^\dag_\uparrow
d^\pdag_\downarrow$, to represent the two states of the low-energy
Hilbert space, the above expression becomes
\begin{eqnarray}
H^{\text{eff}}_\text{int}(T_m)&=&
- i \frac{V^2}{2 \pi a} \int_{0}^{\infty} {\rm d} t\,
{\rm e}^{-i K_z t/2}
\sum_\sigma  \left[
{\rm e}^{-i 2 S_z \sigma (1- W\rho) \phi_\sigma(T_m+t/2) }
{\rm e}^{i 2 S_z \sigma (1- W\rho) \phi_\sigma(T_m-t/2)}
\nonumber \right.\\&& \left.+
\left(
S^+ F^\pdag_\uparrow F^\dag_\downarrow
{\rm e}^{-i \sigma (1- W\rho) \phi_\sigma(T_m+t/2)}
{\rm e}^{i \sigma(1- W\rho) \phi_{-\sigma}(T_m-t/2)}
\nonumber
+ {\rm h.c.}
\right)
\right]\nonumber\,.
\end{eqnarray}
The oscillating factor $e^{-i K_z t/2}$ guarantees that the virtual
excitations are only short lived, and we can therefore expand the term
in the bracket in the small time $t$. Introducing the spin field
$\phi=\frac{1}{\sqrt{2}}\sum_\sigma \sigma \phi_\sigma$, applying the
operator product expansion ${\rm e}^{i \lambda \phi_\sigma(t)} {\rm
  e}^{-i \lambda \phi_\sigma(t')} = (1+i(t-t')/a)^{-\lambda^2}+\lambda
a (1+i(t-t')/a)^{1-\lambda^2} \partial_{t'}\phi_\sigma(t') + \dots$
for the first term, integrating over $t$ using $\int_0^\infty {\rm d}t \, {\rm e}^{-i K_z t/2} (1+i
t/a)^{-\alpha} = -i (a K_z/2)^\alpha 2 \Gamma(1-\alpha)/K_z$ we
obtain in leading order for large $K_z$:
\begin{equation}\label{heff7}
H^{\rm eff}_{\rm int} =
\frac{4 V^2}{K_z \sqrt{2} \pi} (1- W\rho) \Gamma(2-(1- W\rho)^2)
\left(\frac{a K_z}{2}\right)^{(1- W\rho)^2-1}
S_z \,:\partial_{x} \phi(0):
+\frac{4 V^2}{K_z 2 \pi a}\left(
S^+
{\rm e}^{-i \sqrt{2} (1- W\rho) \phi(0)} F^\dag_\downarrow F^\pdag_\uparrow
+ {\rm h.c.}
\right)\,.
\end{equation}
\end{widetext}
Before identifying the coupling constants of the effective low-energy
Hamiltonian two more steps are required. First, we have to re-adjust our
UV cutoff from $a$ to $a_K\sim 1/K_z$ in the definition of our
fields, as we effectively have integrated out short time differences
of order $1/K_z$. To this end we  have to normal-order
$H^\text{eff}_\text{int}$ as only normal-ordered expressions are
cutoff independent,
\bea
e^{i \lambda \phi}
&=&
\left(\frac{2 \pi a}{L}\right)^{\frac{\lambda^2}{2}} :e^{i \lambda \phi}\!: \\
&=&
\left(\frac{a}{a_K}\right)^{\frac{\lambda^2}{2}}
\left(\frac{2 \pi a_K}{L}\right)^{\frac{\lambda^2}{2}} :e^{i \lambda \phi}\!:\;
=\left(\frac{a}{a_K}\right)^{\frac{\lambda^2}{2}}
e^{i \lambda \tilde{\phi}},\nonumber
\eea
where $\tilde{\phi}$ denotes the fields defined with respect to the new cutoff
$a_K$. This effectively leads to the substitution
\begin{equation}
\frac{4 V^2}{K_z 2 \pi a} \to \frac{4 V^2}{K_z 2 \pi a_K} \left(\frac{a}{a_K}\right)^{(1- W\rho)^2-1}
\end{equation}
 in the second term of (\ref{heff7}).

In a last step, we undo the Emery--Kivelson
transformation to obtain the Kondo Hamiltonian in its usual form (\ref{effKondo}) with
\bea\label{jzeff}
J^{\rm eff}_z &=& W + \frac{4 V^2}{K_z} c_W
\left(\frac{a K_z}{2}\right)^{(1- W\rho)^2-1} \,,
\\\label{jperpeff}
J^{\rm eff}_\perp &=& \frac{4 V^2}{K_z} \left(\frac{a}{a_K}\right)^{(1- W\rho)^2-1}\,,
\eea
where $c_W=(1- W\rho) \Gamma(2-(1- W\rho)^2)$.  X-ray edge
singularities induced by $W$ have led to a power-law dependence of the
effective couplings on $K_z$.  Note that the previous arguments fixed
$a_K\sim 1/K_z$ in (\ref{jperpeff}) only up to a prefactor of order
$1$ depending on $W$.
However, this unknown prefactor approaches $1$ close to
the quantum phase transition where $W\rho \to 0$.


\section{Flow equation transformation for the single Kondo impurity}
\label{sect:App_feq}

In this appendix we provide some details on the flow equation treatment
of the single-impurity Kondo model, which was first presented
in Ref.~\onlinecite{HofstetterKehrein_01}.  Here we
will show how to extend the analysis of Ref.~\onlinecite{HofstetterKehrein_01}
to take into account the terms of order $(\lambda^2-1)$ that become
important in our coupled system since they are multiplied by
a possibly large energy scale~$K_z$ in (\ref{H_Aeff}). We refer the reader
to Ref.~\onlinecite{HofstetterKehrein_01} for the basic ideas of the approach
and only present the main steps to keep our presentation here
self-contained.

\subsection{Transformation of the Hamiltonian}

The starting point for the flow equation approach is (\ref{Htilde})
with $\gamma$ chosen such that the longitudinal coupling is eliminated.
This way we arrive at the initial Hamiltonian $H(B=0)$ for the flow
equation approach
\beq
H(B)=H_0[\phi]+\int {\rm d}x\,g(B;x) \left( V(\lambda(B);x) S^- +{\rm h.c.}
\right)
\label{feq1}
\eeq
with $\lambda(B=0)=\lambda_0=\sqrt{2}-J_z/\sqrt{2}\pi v_F$ and
\beq
g(B=0;x)=\delta(x) \left(\frac{2\pi a}{L}\right)^{\lambda(B=0)^2/2}
\frac{J_\perp}{2\pi a} \ .
\eeq
Here $V(\lambda;x)$ are normal ordered vertex operators
\beq
V(\lambda;x)=: e^{-i\lambda\phi(x)} : \ .
\eeq
During the course of the infinitesimal unitary transformations
\beq
\frac{dH(B)}{dB}=[\eta(B),H(B)]
\eeq
with the generator $\eta(B)$ from Ref.~\onlinecite{HofstetterKehrein_01}
\bea
\eta&=&\int {\rm d}x\,\eta^{(1)}(x) \left(V(\lambda;x)S^- -{\rm h.c.}\right) \\
&&+\int {\rm d}x\,{\rm d}x'\,\eta^{(2)}(x,x')\,[V(\lambda;x),V(-\lambda;x')] \nn
\label{feq_eta}
\eea
the interaction $g(B;x)$ in (\ref{feq1}) becomes
more and more nonlocal. With each infinitesimal step of the
transformation one also generates a new interaction term
in (\ref{feq1}) with the structure
\beq
v_F\,S^z \int {\rm d}x\,s(x)\:\partial_x\phi(x)
\label{feq2}
\eeq
and a nonlocal function $s(x)$ that depends on the couplings.
The key step in  Ref.~\onlinecite{HofstetterKehrein_01} is that (\ref{feq2})
can again be eliminated by a unitary transformation of the
Emery--Kivelson type
\beq
U=:\exp\left( i S^z \int {\rm d}x\,s(x)\:\phi(x) \right): \ .
\eeq
\begin{widetext}
We now analyze how the interaction term in (\ref{feq1})
is transformed due to~$U$, e.g.
\bea
U\,V(\lambda;y)S^- U^\dag = :\exp\left( -\frac{i}{2} \int {\rm d}x\,s(x)\:\phi(x) \right):
\: : e^{-i\lambda\phi(y)} : \: :\exp\left( -\frac{i}{2} \int {\rm d}x\,s(x)\:\phi(x) \right):\,S^- \ .
\label{feq4}
\eea
In order to proceed we normal order all the exponentials,
which can be done exactly since the commutator of the bosonic field
is a c-number. This leads to
\bea
U\,V(\lambda;y)S^- U^\dag \propto
:\exp\Big(\sqrt{\frac{2\pi}{L}} \sum_{k>0} \frac{{\rm e}^{-k a/2}}{\sqrt{k}} \big(
 [\lambda {\rm e}^{- i k y}+s(k)]\:  b^\pdag_{k} - [\lambda {\rm e}^{i k y}+s(k)]\:  b^\dag_{k}\big)
\Big): \; S^-
\label{feq3}
\eea
with $s(k)$ being the Fourier transform of $s(x)$ from (\ref{feq2}).
The proportionality factor in (\ref{feq3}) leads to the non-perturbative
renormalization of the coupling constant $g(B;x)$ already obtained
in Ref.~\onlinecite{HofstetterKehrein_01}. Except for the local
coupling at the beginning of the flow the exponential in (\ref{feq3})
cannot be exactly rewritten as a vertex operator. We use two
approximations that give us the correct result up to quadratic
terms in the deviation from the Toulouse line:
i)~We use the infrared limit $s(0)$ instead of $s(k)$
in (\ref{feq3}).
ii)~We expand the exponential in a way that avoids IR-divergences
and neglect higher
order terms in the bosonic operators that lead to irrelevant
couplings:
\bea
\lefteqn{:\exp\Big(\sqrt{\frac{2\pi}{L}} \sum_{k>0} \frac{{\rm e}^{-k a/2}}{\sqrt{k}} \big(
 [\lambda {\rm e}^{- i k y}+s(0)]\:  b^\pdag_{k} -{\rm h.c.} \big)
\Big):} \nn \\
 &=&\quad
:\exp\Big(\sqrt{\frac{2\pi}{L}} \sum_{k>0} \frac{{\rm e}^{-k a/2}}{\sqrt{k}} \big(
 [(\lambda+s(0)) {\rm e}^{- i k y}+(1-e^{-iky}) s(0)]\:  b^\pdag_{k} -{\rm h.c.} \big)
\Big):  \nn \\
&=&\quad
:V(\lambda+s(0);y) \Big( 1+ \sqrt{\frac{2\pi}{L}} \sum_{k>0} \frac{{\rm e}^{-k a/2}}{\sqrt{k}}
\big( (1-e^{-iky}) s(0)\:  b^\pdag_{k} -{\rm h.c.} \big) +\ldots \Big): \,.
\eea
\end{widetext}
Retaining only the first term on the right-hand side is the approximation used
in Ref.~\onlinecite{HofstetterKehrein_01}: one obtains vertex operators
with flowing scaling dimensions $[\lambda+s(0)]$ that eventually become fermionic.
The second term can be understood
as a correction term to this leading behavior due to the nonlocality of
the interaction during the flow equation procedure. It is this term that
eventually leads to the correction term in (\ref{corr_trfsigmaz}).

The above procedure following from (\ref{feq4}) has to be
repeated iteratively throughout the flow, leading to
\bea
\lefteqn{:V(1;y) \Big( 1
+ (1-\lambda_0) \sqrt{\frac{2\pi}{L}} \sum_{k>0}
\frac{{\rm e}^{-k  a_{T_K}/2}}{\sqrt{k}} } \nn \\
&&\qquad\qquad \times\big( (1-e^{-iky})\:  b^\pdag_{k} -{\rm h.c.} \big)
\Big): S^- \nn \\
&=&: \Psi^\dag(y) \Big( 1+(1-\lambda_0)
(-i\bar\phi(0)+i\bar\phi(y))\Big) : S^- \nn \\
&=&: \Psi^\dag(y) \Big( 1+(1-\lambda_0)\, iy\partial_y\bar\phi(0)\Big)
 : S^-
\eea
plus irrelevant terms with higher order derivatives of the
bosonic field. Here
$\bar\phi(y)$ denotes the bosonic spin-density field $\phi(y)$
without the Fourier components for energies larger than
${\cal O}(T_K)$ since the term proportional to $(1-\lambda_0)$
is generated successively during the flow equation procedure.
Putting everything together, the Hamiltonian $H(B)$
from (\ref{feq1}) acquires a new term of order $(1-\lambda_0)$
during the flow that has been neglected in
Ref.~\onlinecite{HofstetterKehrein_01}.
It can be viewed as an assisted hopping term that is
marginal as opposed to the leading order hopping term that
is a relevant operator. The new term can be eliminated by
including an additional term with the structure
\beq
\int {\rm d}x\: \eta^{(3)}(x)\Big(
:V(\lambda;x) \partial_x\bar\phi(0) : S^-
-{\rm h.c.}\Big)
\label{feq_eta3}
\eeq
and a suitable coefficient function $\eta^{(3)}(x)$
into the generator (\ref{feq_eta}). One can verify that
this does not modify the previous flow equations for the
Hamiltonian in linear order in $ \eta^{(3)}(x)$ (essentially
since the assisted hopping term is marginal as opposed to the
relevant hopping term that generates the flow equations in
leading order). Therefore we can neglect these extra terms
in the flow of the Hamiltonian when we want to retain terms
up to linear order in $(\lambda_0-1)$.

\bigskip

\subsection{Transformation of the impurity spin operator}

However, for the transformation of $S^z$ one needs to be more careful
since $S^z$ can be multiplied by a large exchange field $h=\pm K_z/2$
due to the coupling to the second spin. This can be much larger than
the Kondo scale close to the transition.  In order to study the
transformation of $S^z$ we follow the same route as in
Ref.~\onlinecite{HofstetterKehrein_01} by using the identity
$S^z=[S^+,S^-]/2$ and evaluating the transformed $\tilde S^+$ ($\tilde
S^-$ then follows as its hermitean conjugate). One needs to study the
additional effect of (\ref{feq_eta3}) on $S^+$ and finds the following
expression in the low-energy limit: \beq \tilde S^+=\sigma_z \int {\rm d}y\,
d(y)\, : \Psi^\dag(y) \left( 1+(1-\lambda_0)
  i\partial_y\bar\phi(0)\right) : \eeq with [to linear order in
$(\lambda_0-1)$] the same coefficients $d(y)$ as in
Ref.~\onlinecite{HofstetterKehrein_01}. This leads to
\begin{widetext}
\bea
\tilde S^z&=&[\tilde S^+,\tilde S^-]/2 =\frac{1}{2}
\int {\rm d}x\,{\rm d}x'\: d(x)\,d^*(x')\: [\Psi^\dag(x),\Psi(x')]
+\frac{1}{2} (1-\lambda_0) \int   {\rm d}x\,{\rm d}x'\: d(x)\,d^*(x')\:
\nn \\
&&\times\big(-ix'
[\Psi^\dag(x), :\Psi(x')\partial_{x'}\bar\phi(0):] +ix
[:\Psi^\dag(x)\partial_x\bar\phi(0):,\Psi(x')] \big)
 + {\cal O}((\lambda_0-1)^2) \ .
\label{tildeSz}
\eea
Since we are interested in an analysis in the vicinity of the
Toulouse line we only keep terms up to linear order in $(\lambda_0-1)$.
The term of order $(\lambda_0-1)$ consists of two fermionic operators
and a spatial derivative of the bosonic field.
If we subtract the contractions with respect to the ground state
the remaining normal ordered operator will therefore lead to an
irrelevant coupling in the coupled Hamiltonians (\ref{tildeHK}).
However, we need to retain the contractions:
\bea
\tilde S^z&=&\frac{1}{2}
\int {\rm d}x\,{\rm d}x'\: d(x)\,d^*(x')\: [\Psi^\dag(x),\Psi(x')]+\frac{1}{2} (1-\lambda_0) i\partial_x\bar\phi(0)\int  {\rm d}x\,{\rm d}x'\: d(x)\,d^*(x')\:(x-x')
\langle [\Psi^\dag(x),\Psi(x')] \rangle \nn \\
&&+ {\cal O}((\lambda_0-1)^2)+{\rm irrelevant} \nn \\
&=&\frac{1}{2}
\int {\rm d}x\,{\rm d}x'\: d(x)\,d^*(x')\: [\Psi^\dag(x),\Psi(x')]+\frac{1}{2} (\lambda_0-1) f(h)\, \partial_x\bar\phi(0)+ {\cal O}((\lambda_0-1)^2)+{\rm irrelevant}
\eea
with
\beq
f(h)\stackrel{\rm def}{=}
\int {\rm d}k\,{\rm d}k'\:
\left(\partial_k d_k d_{k'}+d_k\partial_{k'} d_{k'} \right)
\langle [\Psi^\dag_k,\Psi_{k'}] \rangle \ .
\eeq
Here $d_k$ denotes the Fourier transform of $d(x)$.
The expectation value $\langle [\Psi^\dag_k,\Psi_{k'}] \rangle$
has to be evaluated in the ground state of the Hamiltonian $H^{(\rm pot)}$
that is obtained from the resonant level
model Hamiltonian plus magnetic field $h\,S^z$ after
the above unitary transformation. Since $S^z$ ``decays''
into fermion operators under this transformation according
to (\ref{tildeSz}), this Hamiltonian is given as
\beq
H^{(\rm pot)}=\sum_k \epsilon_k \Psi^\dag_k \Psi^\pdag_{k}
+\sum_{k,k'} h\,d_k\,d_{k'}\:\Psi^\dag_k\Psi^\pdag_{k'} \ ,
\eeq
i.e.\ this is just a potential scattering model with a
separable potential $V_{kk'}=h\,d_k d_{k'}$. The retarded
Green's function can be calculated in closed form
\beq
G_{kk'}(\epsilon^+)=\frac{\delta_{kk'}}{\epsilon^+-\epsilon_k}
+\frac{h\,d_k d_{k'}}{(\epsilon^+-\epsilon_k)(\epsilon^+-\epsilon_{k'})}\,
\frac{1}{1-\sum_q \frac{h\,d_q^2}{\epsilon^+-\epsilon_q}}
\eeq
leading to
\bea
f(h)&=& \int {\rm d}k\,{\rm d}k'\:
\left(\partial_k d_k d_{k'}+d_k\partial_{k'} d_{k'} \right)
\langle [\Psi^\dag_k,\Psi_{k'}] \rangle  \nn \\
&=&-\frac{1}{\pi} {\rm Im}  \int {\rm d}k\,{\rm d}k'\:
\left(\partial_k d_k d_{k'}+d_k\partial_{k'} d_{k'} \right)\left( \int_{-\infty}^0 {\rm d}\epsilon\,G_{kk'}(\epsilon^+)
- \int_0^{\infty} {\rm d}\epsilon\,G_{kk'}(\epsilon^+)\right) \nn \\
&=&-\frac{1}{\pi} {\rm Im}\left(
\int_{-\infty}^0 {\rm d}\epsilon - \int_0^{\infty} {\rm d}\epsilon \right)  \int {\rm d}k\,
\frac{\partial_k d_k^2}{\epsilon^+-\epsilon_k} \nn \\
&&-h\, \frac{1}{\pi} {\rm Im}\left(
\int_{-\infty}^0 {\rm d}\epsilon - \int_0^{\infty} {\rm d}\epsilon \right)
\int {\rm d}k\,\frac{\partial_k d_k^2}{\epsilon^+-\epsilon_k}\int {\rm d}k'\,\frac{d_{k'}^2}{\epsilon^+-\epsilon_{k'}}
\; \frac{1}{1-\sum_q \frac{h\,d_q^2}{\epsilon^+-\epsilon_q}} \ .
\label{eq:expvalue}
\eea
\end{widetext}
One easily shows that the impurity orbital Green's function
$G_{dd}^{(\epsilon_d=0)}(\epsilon^+)$ in the resonant level model
\beq
H^{(RLM)}=\sum_k \epsilon_k \Psi^\dag_k \Psi^\pdag_{k}
+\sum_k \tilde V (d^\dag \Psi^\pdag_k+ \Psi^\dag_k d)
\eeq
is given by
\beq
G_{dd}^{(\epsilon_d=0)}(\epsilon^+)
=\sum_q \frac{d_q^2}{\epsilon^+-\epsilon_q} \ .
\eeq
Using
\beq
\sum_q \frac{\partial_q d_q^2}{\epsilon^+-\epsilon_q}
=v_F\, \partial_\epsilon G_{dd}^{(\epsilon_d=0)}(\epsilon^+)
\eeq
one can reexpress (\ref{eq:expvalue}) as
\bea
f(h)&=& 2v_F\rho_d^{(\epsilon_d=0)}(0) \nn \\
&&-v_F\,h\, \frac{1}{\pi} {\rm Im}\left(
\int_{-\infty}^0 {\rm d}\epsilon - \int_0^{\infty} {\rm d}\epsilon \right)
\left(\partial_\epsilon  G_{dd}^{(\epsilon_d=0)}(\epsilon^+)\right)
\nn \\
&&\times
G_{dd}^{(\epsilon_d=0)}(\epsilon^+)
\frac{1}{1-h G_{dd}^{(\epsilon_d=0)}(\epsilon^+)} \ ,
\eea
where $\rho_d^{(\epsilon_d=0)}(\epsilon)$ is the impurity orbital
density of states. One notices that the impurity orbital Green's
function in the resonant level model with nonvanishing impurity
orbital energy $\epsilon_d\,d^\dag d$ can be written as
\beq
G_{dd}^{(\epsilon_d)}(\epsilon^+)
=\frac{G_{dd}^{(\epsilon_d=0)}(\epsilon^+)}{
1-\epsilon_d\,G_{dd}^{(\epsilon_d=0)}(\epsilon^+)}
\eeq
which leads to
\bea
f(h)&=& 2v_F\rho_d^{(\epsilon_d=0)}(0) \nn \\
&&-v_F\,h\, \frac{1}{\pi} {\rm Im}\left(
\int_{-\infty}^0 {\rm d}\epsilon - \int_0^{\infty} {\rm d}\epsilon \right) \nn \\
&&~\times
\left(\partial_\epsilon  G_{dd}^{(\epsilon_d=0)}(\epsilon^+)\right)
G_{dd}^{(\epsilon_d=h)}(\epsilon^+) \ .
\eea
This expression can be easily worked out in various limits
\beq
f(h)=\left\{
\begin{array}{ll}
2w\,v_F/T_K^{(1)} & {\rm ~for~} h=0 \\
v_F/|h| & {\rm ~for~} |h|\gg T_K^{(1)}
\end{array}
\right.
\eeq
and a smooth crossover in between (here $w=0.4128$ is the Wilson
number).


\end{document}